\documentclass{revtex4}
\usepackage{graphicx}
\usepackage{latexsym}
\usepackage{amsmath,amssymb}
\usepackage{epsfig,graphicx}
\def\be{\begin{equation}}
\def\ee{\end{equation}}
\def\bea{\begin{eqnarray}}
\def\eea{\end{eqnarray}}

\def\CO{{\cal O}}

\def\bk{{\bf k}}

\def\CA{{\cal A}}

\def\CO{{\cal O}}

\def\Mpl{M_{pl}}

\begin{document}


\title{Non-Gaussianity of quantum fields during inflation }
\author{Kazuya Koyama}
\email{kazuya.koyama.AT.port.ac.uk}%
\affiliation{Institute of Cosmology \& Gravitation, Dennis Sciama Building, University of
Portsmouth, Portsmouth~PO1~3FX, UK}

\begin{abstract}
  In this review, we discuss how non-Gaussianity of cosmological perturbations arises from inflation. After introducing the in-in formalism to calculate the $n$-point correlation function of quantum fields, we present the computation of the bispectrum of the curvature perturbation generated in general single field inflation models. The shapes of the bispectrum are compared with the local-type non-Gaussianity that arises from non-linear dynamics on super-horizon scales.
\end{abstract}

\maketitle

\section{Introduction}
There is currently a great deal of interest in the statistical properties of primordial
perturbations from inflation, because measurements of any non-Gaussianity will improve by
about an order of magnitude over the next few years, for example with the Planck
\cite{PLANCK}. This will provide a key way to
discriminate between the many models of inflation. Although single field models of
slow-roll inflation typically generate a small level of non-Gaussianity
\cite{Acquaviva:2002ud,Maldacena:2002vr}, there may be an observable level generated in
many alternative models of early universe
\cite{Linde:1996gt,Bartolo:2001cw,Bernardeau:2002jy,Bernardeau:2002jf,Dvali:2003em,Creminelli:2003iq, Alishahiha:2004eh,Gruzinov:2004jx,Enqvist:2004ey,Jokinen:2005by,Enqvist:2005qu,Lyth:2005qk,Salem:2005nd,Seery:2006js,
Seery:2005wm, Seery:2005gb, Sasaki:2006kq,Malik:2006pm,Barnaby:2006cq,Alabidi:2006wa,Chen:2006nt,Huang:2006eh,Chen:2006xjb,Alabidi:2006hg,
Byrnes:2006vq,
Arroja:2008ga,Arroja:2008yy,Langlois:2008wt,Langlois:2008qf,
Sasaki:2008uc,Byrnes:2008wi,Byrnes:2008zy,Yokoyama:2007uu, Yokoyama:2007dw, Dutta:2008if,Naruko:2008sq,Suyama:2007bg, Suyama:2008nt, Gao:2008dt,Cogollo:2008bi,Rodriguez:2008hy,Ichikawa:2008iq,Byrnes:2008zz,Li:2008fma,Langlois:2008vk,Hikage:2008sk,
Kawasaki:2008sn,Creminelli:2007aq, Koyama:2007if, Buchbinder:2007at, Lehners:2007wc, Lehners:2008my, Misra:2008tx, Huang:2009vk, Khoury:2008wj}.

We are interested in the primordial curvature perturbation on uniform density hypersurfaces, $\zeta$, on large scales, which is directly related to temperature anisotropies in Cosmic Microwave background (CMB). The power spectrum of $\zeta$ is defined as
\begin{equation}
\langle \zeta(\bk_1) \zeta(\bk_2) \rangle
=(2 \pi)^3 \delta^{(3)}(\bk_1 + \bk_2) P_{\zeta}(k_1),
\end{equation}
where $\zeta(\bk)$ is a Fourier component of $\zeta(t,x^i)$.
If $\zeta$ obeys Gaussian statistics, the power spectrum determines all statistical quantities. However, if the distribution function of $\zeta$ deviates from Gaussian statistics, we need to specify higher order statistics. The first non-trivial statistics is the bispectrum defined by
\begin{equation}
\langle \zeta(\bk_1) \zeta(\bk_2) \zeta(\bk_2) \rangle
=(2 \pi)^3 \delta^{(3)}(\bk_1 + \bk_2 + \bk_3) B_{\zeta}(k_1,k_2,k_3).
\end{equation}
Currently observations of the CMB have concentrated on constraining the
3-point function (bispectrum) \cite{Verde:1999ij,Wang:1999vf,Komatsu:2001rj}.

The most instructive way to understand how non-vanishing bispectrum appears from non-linearities is to use the so-called delta-N formalism \cite{Starobinsky:1986fxa,Sasaki:1995aw,Lyth:2004gb,Lyth:2005fi}.
See \cite{Tanaka:2010km} for a concise review of the delta-N formalism.
This is based on the separate universe approach \cite{Salopek:1990jq,Sasaki:1998ug,Wands:2000dp}.
This considers each super-Hubble scale patch to be evolving like a separate Friedman-Robertson-Walker universe which is locally homogeneous. By patching these regions together we can track the evolution of the curvature perturbation on large scales just by using background quantities.

The number of $e$-foldings, $N$, given by
\bea N=\int^{t_{\rm{fin}}}_{t_{\rm{ini}}}H(t) dt\,, \eea
is evaluated from an initial flat hypersurface to a final uniform-density hypersurface.~The perturbation in the number of $e$-foldings, $\delta N$, is the difference between the curvature perturbations on the initial and final hypersurfaces. We wish to calculate primordial perturbations, hence we pick a final uniform density
hypersurface to be at a fixed time during the standard radiation dominated era, for
example during primordial nucleosynthesis. The initial time is arbitrary provided it is
after the Hubble exit time of all relevant scales. It is often convenient to pick this
time to be shortly after Hubble-exit.

Let us consider a model where a scalar field $\phi$ determines the expansion of the Universe. The scalar field acquires quantum fluctuations $\delta \phi$
under horizon scales which cause fluctuations
in $e$-foldings in each super-Hubble scale patch. Then $\zeta$ can be written as
\begin{equation}
\zeta(t,x^i) = N_{,\phi} (t,t_*) \delta \phi(t_*,x^i) +
\frac{1}{2}
N_{,\phi \phi}(t,t_*) \left(
\delta \phi(t_*,x^i)^2   - \langle \delta \phi(t_*,x^i)^2 \rangle   \right) +...,
\label{deltaN}
\end{equation}
where $t_*$ denotes the horizon crossing time and $N_{,\phi} = d N/d \phi$ and $N_{,\phi \phi}
= d^2 N/d \phi^2$.
Using Eq.~(\ref{deltaN}), the bispectrum of the curvature perturbation is calculated as
\begin{eqnarray}
\label{connect}
  \langle \zeta(\mathbf{k}_1) \zeta(\mathbf{k}_2) \zeta(\mathbf{k}_3) \rangle
  &=& N_{,\phi}^3 \langle \delta \phi(\mathbf{k}_1)
  \delta \phi(\mathbf{k}_2) \delta \phi(\mathbf{k}_3) \rangle
   + \frac{
    N_{,\phi \phi} N_{,\phi}^2}{2}
   \langle \delta \phi(\mathbf{k}_1) \delta \phi(\mathbf{k}_2) [\delta \phi \star
  \delta \phi](\mathbf{k}_3) \rangle + \mbox{perms},
\end{eqnarray}
where $[\delta \phi \star \delta \phi]$ denotes a convolution
\begin{equation}
 [\delta \phi \star \delta \phi]  (\mathbf{k}) = \int \frac{d^3 q}{(2 \pi)^3}
  \delta \phi (\mathbf{q})  \delta \phi (\mathbf{k}-\mathbf{q}).
\end{equation}
A diagrammatic approach to compute the higher order correlation function is developed in Ref.~\cite{Byrnes:2007tm}.

There are in general two contributions to the bispectrum of the curvature perturbations. One arises from the bispectrum of quantum fluctuations of the scalar field generated under horizon scales,
$\langle \delta \phi(\mathbf{k}_1) \delta \phi(\mathbf{k}_2) \delta \phi(\mathbf{k}_3) \rangle$. The other is coming from the non-linear evolution of the scalar field on super-horizon scales determined by the second derivative of $N$ with respect to the field, $N_{,\phi \phi}$. The latter contribution exists even if the field perturbations at the horizon crossing are Gaussian. This contribution is often called the local type non-Gaussianity as this arises from a non-linear local relation between the curvature perturbation $\zeta$ and the field perturbations $\delta \phi$ in real space. On the other hand, the contribution from the non-linearity of quantum fields depends on non-linear interactions under horizon scales and its $k$-dependence is in general very different from the local type non-Gaussianity.  In this review, we derive this contribution in several inflation models and compare its shape with the local type non-Gaussianity.

The structure of this review is as follows. In section II, we review the in-in formalism to calculate the $n$-point function of quantum fields. In section III, the effective action for higher order perturbations is derived in two gauges that are necessary to compute the interaction Hamiltonian at third order. We calculate the bispectrum of the curvature perturbation in slow-roll inflation and k-inflation.
In section V, we compare the shape of the bispectrum in k-inflation models with that in the local-type non-Gaussianity. Section VI is devoted to conclusions.

\section{Quantum correlations in the in-in formalism}
In this section, we review the in-in formalism to calculate the $n$-point function of quantum fields by following Ref.~\cite{Weinberg:2005vy}. Consider a general Hamiltonian system, with canonical variables $\phi({\bf x},t)$ and conjugates $\pi({\bf x},t)$ satisfying the commutation relations
\begin{equation}
\Big[\phi({\bf x},t),\pi({\bf y},t)\Big]=i \delta^{3}({\bf x}-{\bf y})\;,~~~~~\Big[\phi({\bf x},t),\phi({\bf y},t)\Big]= \Big[\pi({\bf x},t),\pi({\bf y},t)\Big]=0\;,
\label{com}
\end{equation}
and the equations of motion
\begin{equation}
\dot{\phi}({\bf x},t)=i\Big[H[\phi(t),\pi(t)],\phi({\bf x},t)\Big] \;,~~~~~~~ \dot{\pi}({\bf x},t)=i\Big[H[\phi(t),\pi(t)],\pi({\bf x},t)\Big] \;.
\label{Heq}
\end{equation}
The Hamiltonian $H$ is a functional of the $\phi({\bf x},t)$ and $\pi({\bf x},t)$ at fixed time $t$, which according to Eq.~(\ref{Heq}) is independent of the time at which these variables are evaluated.

We assume the existence of a time-dependent classical number solution $\phi_0({\bf x},t)$,
$\pi_0({\bf x},t)$, satisfying the classical equations of motion
and we expand around this solution, writing
\begin{equation}
\phi({\bf x},t)=\phi_0(t)+\delta\phi({\bf x},t)\;,~~~~~~~~~~
\pi({\bf x},t)=\pi_0(t)+\delta\pi({\bf x},t)\;.
\end{equation}
Since classical numbers commute with everything, the fluctuations satisfy the same commutation rules
(\ref{com}) as the total variables.

Now, although $H$ generates the time-dependence of $\phi({\bf x},t)$ and $\pi({\bf x},t)$, it is $\tilde{H}$ rather than $H$ that generates the time dependence of $\delta\phi({\bf x},t)$ and $\delta\pi({\bf x},t)$ where $\tilde{H}[\delta\phi(t),\delta\pi(t);t]$ is the sum of all terms in $H[\phi_0(t)+\delta\phi(t), \pi_0(t)+\delta\pi(t)]$ of second and higher order in the $\delta\phi({\bf x},t)$ and/or $\delta\pi({\bf x},t)$:
\begin{equation}
\delta\dot{\phi}({\bf x},t)=i\Big[\tilde{H}[\delta \phi({\bf x}, t),\delta\pi({\bf x}, t);t],\delta\phi({\bf x},t)\Big] \;,~~~~~~~ \delta\dot{\pi}({\bf x},t)=i\Big[\tilde{H}[\delta \phi({\bf x}, t),\delta \pi({\bf x}, t);t],\delta\pi({\bf x},t)\Big] \;.
\label{Heqfluc}
\end{equation}
This then is our prescription for constructing the time-dependent Hamiltonian $\tilde{H}$ that governs the time-dependence of the fluctuations: expand the original Hamiltonian $H$ in powers of fluctuations $\delta\phi$ and $\delta\pi$, and throw away the terms of zeroth and first order in these fluctuations.  It is this construction that gives $\tilde{H}$ an explicit dependence on time. It follows from Eq.~(\ref{Heqfluc}) that the fluctuations at time $t$ can be expressed in terms of the same operators at some very early time $t_0$ through a unitary transformation
\begin{equation}
\delta\phi(t)=U^{-1}(t,t_0)\delta\phi(t_0)\,U(t,t_0)\;,~~~~
\delta\pi(t)=U^{-1}(t,t_0)\delta\pi(t_0)\,U(t,t_0)\;,
\label{sol}
\end{equation}
where $U(t,t_0)$ is defined by the differential equation
\begin{equation}
\frac{d}{dt}U(t,t_0)=-i\,\tilde{H}[\delta\phi(t),\delta\pi(t);t]\,U(t,t_0),
\label{eqU}
\end{equation}
and the initial condition $U(t_0,t_0)=1$.
In the application that concerns us in cosmology, we can take $t_0=-\infty$, by which we mean any time early enough so that the wavelengths of interest are deep inside the horizon.
From now on, we will omit the ${\bf x}$ dependence of $\delta \phi$ and $\delta \pi$ to simplify the notation.

To calculate $U(t,t_0)$, we now further decompose $\tilde{H}$ into a kinematic term $H_0$ that is quadratic in the fluctuations, and an interaction term $H_I$:
\begin{equation}
\tilde{H}[\delta\phi(t),\delta\pi(t);t]=H_0[\delta\phi(t),\delta\pi(t);t]+H_I[\delta\phi(t),
\delta\pi(t);t]\;,
\end{equation}
and we seek to calculate $U$ as a power series in $H_I$.  To this end, we introduce an ``interaction picture'': we define fluctuation operators $\delta\phi^I(t)$ and $\delta\pi^I(t)$ whose time
dependence is generated by the quadratic part of the Hamiltonian:
\begin{equation}
\delta\dot{\phi}^I(t)=i\Big[H_0[\delta\phi^I(t),\delta\pi^I(t);t],\delta\phi^I(t)\Big] \;,~~~~~~~ \delta\dot{\pi}^I(t)=i\Big[H_0[\delta\phi^I(t),\delta\pi^I(t);t],\delta\pi^I(t)\Big] \;,
\label{Heqint}
\end{equation}
and the initial conditions $\delta{\phi}^I(t_0)=\delta{\phi}(t_0),
\delta{\pi}^I(t_0)=\delta{\pi}(t_0)$. Because $H_0$ is quadratic, the interaction picture operators are free fields, satisfying linear wave equations.

It follows from Eq.~(\ref{Heqint}) that in evaluating $H_0[\delta\phi^I,\delta\pi^I;t]$ we can take the time argument of $\delta\phi^I$ and $\delta\pi^I$ to have any value, and in particular we can take it as $t_0$, so that
$H_0[\delta\phi^I(t),\delta\pi^I(t);t]=H_0[\delta\phi(t_0),\delta\pi(t_0);t]\;$,
but the intrinsic time-dependence of $H_0$ still remains.  The solution of Eq.~(\ref{Heqint})
can again be written as a unitary transformation:
\begin{equation}
\delta\phi_a^I(t)=U^{-1}_0(t,t_0)\delta\phi_a(t_0)U_0(t,t_0)\;,~~~~
\delta\pi_a^I(t)=U^{-1}_0(t,t_0)\delta\pi_a(t_0)U_0(t,t_0)\;,
\end{equation}
with $U_0$ defined by the differential equation
\begin{equation}
\frac{d}{dt}U_0(t,t_0)=-i\,H_0[\delta\phi(t_0),\delta\pi(t_0);t]\,U_0(t,t_0)
\label{eqU0}
\end{equation}
and the initial condition $U_0(t_0,t_0)=1\;$.
Then from Eqs.~(\ref{eqU}) and (\ref{eqU0}) we have
$$
\frac{d}{dt}\Big[U_0^{-1}(t,t_0)U(t,t_0)\Big]=-iU_0^{-1}(t,t_0)H_I[\delta\phi(t_0),\delta\pi(t_0);t]U(t,t_0)\;.
$$
This gives
\begin{equation}
U(t,t_0)=U_0(t,t_0)F(t,t_0)\;, \quad
\frac{d}{dt}F(t,t_0)=-iH_I(t)F(t,t_0)\;,~~~~F(t_0,t_0)=1\;.
\label{eqF}
\end{equation}
where $H_I(t)$ is the interaction Hamiltonian in the interaction picture:
\begin{equation}
H_I(t)\equiv U_0(t,t_0)H_I[\delta\phi(t_0),\delta\pi(t_0);t]U_0^{-1}(t,t_0)=H_I[\delta\phi^I(t),\delta\pi^I(t);t]
\end{equation}
The solution of equations like (\ref{eqF}) is well known (see for example \cite{Peskin:1995ev})
\begin{equation}
F(t,t_0)=T\exp\left(-i\int_{t_0}^t H_I(t)\,dt\right),
\label{solF}
\end{equation}
where $T$ indicates that the products of $H_I$s in the power series expansion of the exponential are to be time-ordered; that is, they are to be written from left to right in the decreasing order of time arguments.  The solution for the fluctuations in terms of the free fields of the interaction picture is given by Eqs.~(\ref{sol}) and (\ref{solF}). Then expectation values of
some product ${\cal A}(t)$ of field operators are obtained as
\begin{equation}
\langle {\cal A}(t) \rangle
=\left \langle \left[\bar{T}\exp\left(i\int_{t_0}^t H_I(t)\,dt\right)\right]\,
{\cal A}^I(t)\,\left[T\exp\left(-i\int_{t_0}^t H_I(t)\,dt\right)\right]
\right \rangle
\;,
\label{A}
\end{equation}
where ${\cal A}(t)$ is any $\delta\phi({\bf x},t)$ or $\delta\pi({\bf x},t)$ or any product of the $\delta\phi$s
and/or $\delta\pi$s, all at the same time $t$ but in general with different space coordinates, and
${\cal A}^I(t)$ is the same product of $\delta\phi^I({\bf x},t)$ and/or $\delta\pi^I({\bf x},t)$.  Also, $\bar{T}$ denotes anti-time-ordering: products of $H_I$s in the power series expansion of the exponential  are to be written from left to right in the increasing order of time arguments.
It is more convenient to use a formula equivalent to Eq.~(\ref{A}):
\begin{equation}
\langle {\cal A}(t)\rangle = \sum_{N=0}^\infty i^N\, \int_{-\infty}^t dt_N \int_{-\infty}^{t_N} dt_{N-1} \cdots \int_{-\infty}^{t_2} dt_1 \left\langle \Big[H_I(t_1),\Big[H_I(t_2),\cdots \Big[H_I(t_N),{\cal A}^I(t)\Big]\cdots\Big]\Big]\right\rangle\;.
\label{in-in}
\end{equation}
In the following section, we apply this formula to calculate the bispectrum of quantum fields generated during inflation.


\section{Non-linear cosmological perturbations}
In this section, we calculate the action for higher order cosmological perturbations that is necessary to compute the interaction Hamiltonian at the third order in the in-in formalism. This calculation is pioneered by
Ref.~\cite{Maldacena:2002vr} and extended to general inflation models by Ref.~\cite{Seery:2005wm, Seery:2005gb, Chen:2006nt}. Here we review the derivation of the higher order action by following Ref.~\cite{Chen:2006nt, Arroja:2008ga}.

\subsection{Inflation models}
To set up our notation, let us first review the formalism in
\cite{Garriga:1999vw} where a general Lagrangian for the inflaton field
is considered. The Lagrangian is of the general form
\begin{equation} \label{generalaction}
S=\frac{1}{2}\int d^4x \sqrt{-g} \left[M_{pl}^2R +
2P(X,\phi)\right]~,
\end{equation}
where $\phi$ is the inflaton field and
$X=-(1/2) g^{\mu\nu}\partial_{\mu}\phi \partial_{\nu}\phi$.
The reduced Planck mass is $M_{pl}=(8\pi G)^{-1/2}$ and the
signature of the metric is $(-1,1,1,1)$.
The energy of the inflaton field is
\begin{equation}
E=2X P_{,X} -P ~,
\label{EPdef}
\end{equation}
where $P_{,X}$ denote the derivative with respect to $X$.
Suppose the universe is homogeneous with a
Friedmann-Robertson-Walker metric
\begin{equation}
ds^2=-dt^2+a^2(t)dx_3^2 ~.
\end{equation}
Here $a(t)$ is the scale factor and $H=\dot{a}/a$ is the
Hubble parameter of the universe. The equations of motion of the
gravitational dynamics are the Friedmann equation and the
continuity equation
\begin{equation}
3M_{pl}^2H^2 = E \label{eomH} ~, \quad
\dot{E} =-3H(E+P) ~.
\end{equation}
It is useful to define the ``speed of sound'' $c_s$ as (see \cite{Christopherson:2008ry, Arroja:2010wy} for a definition
of the sound speed)
\begin{eqnarray}
c_s^2 = \frac{P_{,X}}{E_{,X}}= \frac{P_{,X}}{P_{,X}+2X P_{,XX}},
\end{eqnarray}
and some ``slow variation parameters'' as
in standard slow roll inflation
\begin{equation} \label{small}
\epsilon = -\frac{\dot{H}}{H^2}=\frac{X P_{,X}}{\Mpl^2 H^2}~, \quad
\eta = \frac{\dot{\epsilon}}{\epsilon H}~, \quad
s = \frac{\dot{c_s}}{c_s H} ~.
\end{equation}
These parameters are more general than the usual slow roll
parameters (which are defined through properties of a flat potential,
assuming canonical kinetic terms),
and in general
depend on derivative terms as well as the potential.
For example, in Dirac-Born-Infeld (DBI)
inflation the potential can be steep, and kinetically driven
inflation can occur even in absence of a potential.
We also note that the
smallness of the parameters $\epsilon$, $\eta$, $s$ does not
imply that the rolling of inflaton is slow. When we refer to the slow-roll
expansion, we assume that all the three slow variation parameters are small.

The primordial power spectrum is derived for this general
Lagrangian in \cite{Garriga:1999vw}
\begin{equation} \label{power1}
{\cal P}_{\zeta}(k)=\frac{1}{36 \pi^2
M_{pl}^4}\frac{E^2}{c_s(P+E)}=\frac{1}{8 \pi^2
M_{pl}^2}\frac{H^2}{c_s\epsilon} ~,
\end{equation}
where the expression is evaluated at the time of horizon exit at
$c_s k=aH$ and ${\cal P}_{\zeta}(k) = k^3 P_{\zeta}(k)/2 \pi^2$.
The spectral index is
\begin{equation} \label{index1}
n_s-1=\frac{d\ln {\cal P}_{\zeta}(k)}{d \ln k}= -2\epsilon-\eta-s ~.
\end{equation}
In order to have an almost scale invariant power spectrum, we need
to require the 3 parameters $\epsilon$, $\eta$, $s$ to be very
small, which we will denote simply as $\CO(\epsilon)$.
We note that in inflationary models with
standard kinetic terms the speed of sound is $c_s=1$.
In the case of DBI
inflation, the speed of sound can be very small.
In the case of arbitrary $c_s$, Eqs.~(\ref{power1}) and (\ref{index1})
for the power spectrum and its index at
leading order is still valid as long as the variation of the sound
speed is slow, namely $s\ll 1$. In the following we set $M_{pl}=1$.

\subsection{Effective action for higher order perturbations}
Now in the general setup described by the action (\ref{generalaction}), we need to
expand the action up to the cubic order in perturbations to obtain the third order interacting Hamiltonian.
For this purpose,it is convenient to use the ADM metric formalism \cite{Arnowitt:1960es}.
The ADM line element reads
\begin{equation}
ds^2=-N^2dt^2+h_{ij}\left(dx^i+N^idt\right)\left(dx^j+N^jdt\right),
\label{ADMmetricphi}
\end{equation}
where $N$ is the lapse function, $N^i$ is the shift vector and
$h_{ij}$ is the 3D metric.

The action (\ref{generalaction}) becomes
\begin{equation}
S=\frac{1}{2}\int dtd^3x\sqrt{h}N\left({}^{(3)}\!R+2P\right)+
\frac{1}{2}\int dtd^3x\sqrt{h}N^{-1}\left(E_{ij}E^{ij}-E^2\right).
\end{equation}
The tensor $E_{ij}$ is defined as
\begin{equation}
E_{ij}=\frac{1}{2}\left(\dot{h}_{ij}-\nabla_iN_j-\nabla_jN_i\right),
\end{equation}
and it is related to the extrinsic curvature by
$K_{ij}=N^{-1}E_{ij}$. $\nabla_i$ is the covariant derivative with
respect to $h_{ij}$ and all contra-variant indices in this section
are raised with $h^{ij}$ unless stated otherwise.

The Hamiltonian and momentum constraints are respectively
\begin{eqnarray}
{}^{(3)}\!R+2P-2\pi^2N^{-2}P_{,X}-N^{-2}\left(E_{ij}E^{ij}-E^2\right)&=&0,\nonumber\\
\nabla_j\left(N^{-1}E_i^j\right)-\nabla_i\left(N^{-1}E\right)&=&\pi
N^{-1}\nabla_i\phi P_{,X},\label{LMphi}
\end{eqnarray}
where $\pi$ is defined as
\begin{equation}
\pi\equiv \dot{\phi}-N^j\nabla_j\phi.\label{pi}
\end{equation}
We decompose the shift vector $N^i$ into scalar and intrinsic
vector parts as
\begin{equation}
N_i=\tilde{N_i}+\partial_i\psi,
\end{equation}
where $\partial_i\tilde{N^i}=0$, here indices are raised with
$\delta_{ij}$.

Before we consider perturbations around our background let us
count the number of degrees of freedom (dof) that we have. There
are five scalar functions, the field $\phi$, $N$, $\psi$,
$\mbox{det} h$ and $h_{ij}\sim\partial_i\partial_j H$, where $H$
is a scalar function and $\mbox{det} h$ denotes the determinant of
the 3D metric. Also, there are two vector modes $\tilde{N}^i$ and
$h_{ij}\sim\partial_i\chi_j$, where $\chi^j$ is an arbitrary
vector. Both $\tilde{N}^i$ and $\chi^j$ satisfy a divergenceless
condition and so carry four dof. Furthermore, we also have a
transverse and traceless tensor mode $\gamma_{ij}$ that contains
two additional dof. Because our theory is invariant under change
of coordinates we can eliminate some of these dof. For instance, a
spatial reparametrization like
$x^i=\tilde{x}^i+\partial^i\tilde{\epsilon}(\tilde{x},\tilde{t})
+\epsilon^i_{(t)}(\tilde{x},\tilde{t})$,
where $\tilde{\epsilon}$ and  $\epsilon^i_{(t)}$ are arbitrary and
$\partial_i\epsilon^i_{(t)}=0$, can be chosen so that it removes
one scalar dof and one vector mode. A time reparametrization would
eliminate another scalar dof. Constraints in the action will
eliminate further two scalar dof and a vector mode. In the end we
are left with one scalar, zero vector and one tensor modes that
correspond to three physical propagating dof. In this review, we are
primarily interested in a scalar degree of freedom.

In order to identify this scalar degree of freedom, we need to fix
a gauge. There are two commonly used gauges. In the next subsection
we derive the higher order action in these gauges.

\subsection{Non-linear perturbations in the comoving gauge}

In the comoving gauge, the scalar degree of freedom is the so-called curvature
perturbation $\zeta$ and the inflaton fluctuations
vanish. The 3D metric is perturbed as
\begin{eqnarray}
&&\delta\phi=0,\nonumber\\  &&h_{ij}=a^2e^{2\zeta}\hat{h}_{ij},
\quad
\hat{h}_{ij}=\delta_{ij}+\gamma_{ij}+\frac{1}{2}\gamma_{ik}\gamma_j^k+\cdots\label{zetagauge}
\end{eqnarray}
where $\mbox{det} \hat{h}=1$, $\gamma_{ij}$ is a tensor
perturbation that we assume to be a second order quantity, i. e.
$\gamma_{ij}=\mathcal{O}(\zeta^2)$. It obeys the traceless and
transverse conditions $\gamma_i^i=\partial^i\gamma_{ij}=0$
(indices are raised with $\delta_{ij}$). $\zeta$ is the gauge
invariant scalar perturbation. In (\ref{zetagauge}), we have
ignored the first order tensor perturbations
${}^{(1)}{\gamma_{ij}}_{GW}$. This is because any correlation
function involving this tensor mode will be smaller than a
correlation function involving only scalars, see results of
\cite{Maldacena:2002vr}.

We expand $N$ and $N^i$ in power of the perturbation $\zeta$
\begin{eqnarray}
N=1+\alpha_1+\alpha_2+\cdots,\\
\tilde{N_i}=\tilde{N_i}^{(1)}+\tilde{N_i}^{(2)}+\cdots,\\
\psi=\psi_1+\psi_2+\cdots,
\end{eqnarray}
where $\alpha_n$, $\tilde{N_i}^{(n)}$ and $\psi_n$ are of order
$\zeta^n$.
In order to compute the effective action to order $\CO(\zeta^3)$,
as pointed out in \cite{Maldacena:2002vr}, in the ADM
formalism one only needs to consider the
perturbations of $N$ and $N^i$ to the first order $\CO(\zeta)$. This
is because their perturbations at order
$\CO(\zeta^{3})$ such as $\alpha_3$ will multiply the
constraint equation
at the zeroth order $\CO(\zeta^0)$ which vanishes, and the second
order perturbations such as $\alpha_2$ will
multiply a factor which vanishes by the first
order solution. So the first order solution for
$N$ and $N^i$ is enough for our purpose.
Therefore our task is simplified. In order to expand the action
(\ref{generalaction}) to
quadratic and cubic order in the
primordial scalar perturbation $\zeta$, we only need to plug in the
solution for the first
order perturbation in $N$ and $N^i$ and do the expansion.

Now, the strategy is to solve the constraint equations for the
lapse function and shift vector in terms of $\zeta$ and then plug
in the solutions in the expanded action up to third order.
At first order in $\zeta$, a particular solution for equations
(\ref{LMphi}) is \cite{Maldacena:2002vr,Seery:2005wm}:
\begin{equation}
\alpha_1=\frac{\dot{\zeta}}{H}, \quad \tilde{N_i}^{(1)}=0, \quad
\psi_1=-\frac{\zeta}{H}+\chi, \quad
\partial^2\chi=a^2\frac{\epsilon}{c_s^2}\dot{\zeta}.\label{N1order}
\end{equation}

The second order action is
\begin{equation}
S_2=\int
dtd^3x\left[a^3\frac{\epsilon}{c_s^2}\dot{\zeta}^2-a\epsilon\left(\partial\zeta\right)^2\right].
\label{2actionz}
\end{equation}
The third order action is
\cite{Maldacena:2002vr,Seery:2005wm,Chen:2006nt}
\begin{eqnarray}
S_3&=&\int dtd^3x\left[-\epsilon
a\zeta\left(\partial\zeta\right)^2-a^3\left(\Sigma+2\lambda\right)\frac{\dot{\zeta}^3}{H^3}
+\frac{3a^3\epsilon}{c_s^2}\zeta\dot{\zeta}^2\right.\nonumber\\
&&\left.+\frac{1}{2a}\left(3\zeta-\frac{\dot{\zeta}}{H}\right)\left(\partial_i\partial_j\psi_1\partial_i\partial_j\psi_1-\partial^2\psi_1\partial^2\psi_1\right)
-\frac{2}{a}\partial_i\psi_1\partial_i\zeta\partial^2\psi_1\right].\label{3actionz}
\end{eqnarray}
Here we defined two parameters following \cite{Seery:2005wm}
\begin{eqnarray}
\Sigma&=&X P_{,X}+2X^2P_{,XX}  = \frac{H^2\epsilon}{c_s^2} ~,\\
\lambda&=& X^2P_{,XX}+\frac{2}{3}X^3P_{,XXX} ~.
\label{lambda}
\end{eqnarray}


\subsection{\label{subsec:PerturbationsUniformeR}Non-linear
perturbations in the uniform curvature gauge}
In this gauge, the inflaton perturbation does not vanish and the
3D metric takes the form
\begin{eqnarray}
&&\phi(x,t)=\phi_0+\delta\phi(x,t),\nonumber\\
&&h_{ij}=a^2\hat{h}_{ij}, \quad
\hat{h}_{ij}=\delta_{ij}+\tilde{\gamma}_{ij}+\frac{1}{2}\tilde{\gamma}_{ik}
\tilde{\gamma}_j^k+\cdots\label{deltaphigauge}
\end{eqnarray}
where $\mbox{det} \hat{h}=1$ and $\tilde{\gamma}_{ij}$ is a tensor
perturbation that we assume to be a second order quantity, i.e.,
$\tilde{\gamma}_{ij}=\mathcal{O}(\delta\phi^2)$. It obeys the
traceless and transverse conditions
$\tilde{\gamma}_i^i=\partial^i\tilde{\gamma}_{ij}=0$ (indices are
raised with $\delta_{ij}$).

We expand $N$ and $N^i$ in powers of the perturbation
$\delta\phi(x,t)$
\begin{eqnarray}
N=1+\alpha_1+\alpha_2+\cdots,\\
\tilde{N_i}=\tilde{N_i}^{(1)}+\tilde{N_i}^{(2)}+\cdots,\\
\psi=\psi_1+\psi_2+\cdots,
\end{eqnarray}
where $\alpha_n$, $\tilde{N_i}^{(n)}$ and $\psi_n$ are of order
$\delta\phi^n$ and $\phi_0(t)$ is the background value of the
field. At first order in $\delta\phi$, a particular solution for
equations (\ref{LMphi}) is \cite{Maldacena:2002vr,Seery:2006vu}:
\begin{equation}
\alpha_1=\frac{1}{2H}\dot{\phi_0}\delta\phi P_{,X}, \quad
\tilde{N_i}^{(1)}=0, \quad
\partial^2\psi_1=\frac{a^2\epsilon}{c_s^2}\frac{d}{dt}\left(-\frac{H}{\dot{\phi}}\delta\phi\right)
. \label{N1orderphi}
\end{equation}

The second-order action is given by
\begin{eqnarray}
S_2=\int dtd^3xa^3&\!\!\!\!\Big[&\!\!\!\!
P_{,XX}X_0\left(\dot{\delta\phi}^2+2X_0\alpha_1^2-2\dot\phi_0\alpha_1\dot{\delta\phi}\right)
+P_{,X\phi}\left(\dot\phi_0\delta\phi\dot{\delta\phi}-2X_0\alpha_1\delta\phi\right)
+\frac{1}{2}P_{,\phi\phi}{\delta\phi}^2
\nonumber\\
&&+P_{,X}\left(\frac{1}{2}\dot{\delta\phi}^2-\dot\phi_0\alpha_1\dot{\delta\phi}+X_0\alpha_1^2
             -a^{-2}\left(\frac{1}{2}(\partial\delta\phi)^2+\dot\phi_0\partial_i\delta\phi\partial^i\psi_1\right)
      \right)
\nonumber\\
&&-3H^2\alpha_1^2+P_{,\phi}\alpha_1\delta\phi-2a^{-2}H\alpha_1\partial^2\psi_1
\Big],\label{2actionucurvature}
\end{eqnarray}
where $X_0=\dot\phi_0^2/2$.
The third-order action is obtained as
\begin{eqnarray}
\label{3actiono}
S_3=\int dtd^3xa^3&\!\!\!\!\Bigg[&\!\!\!\!
P_{,XX}\Bigg(\frac{1}{2}\dot\phi_0\dot{\delta\phi}^3+X_0\alpha_1\left(-4\dot{\delta\phi}^2+5\dot\phi_0\alpha_1\dot{\delta\phi}-4X_0\alpha_1^2\right)
             \nonumber\\
             &&\qquad\quad+a^{-2}\left(-\frac{1}{2}\dot\phi_0\dot{\delta\phi}(\partial\delta\phi)^2+X_0\alpha_1(\partial\delta\phi)^2-2X_0\left(\dot{\delta\phi}-\dot\phi_0\alpha_1\right)\partial_i\delta\phi\partial^i\psi_1\right)
        \Bigg)
\nonumber\\
&&+P_{,X\phi}\Bigg(\frac{1}{2}\delta\phi\dot{\delta\phi}^2-\dot\phi_0\alpha_1\delta\phi\dot{\delta\phi}+X_0\alpha_1^2\delta\phi
                   -a^{-2}\left(\frac{1}{2}\delta\phi(\partial\delta\phi)^2+\dot\phi_0\delta\phi\partial_i\delta\phi\partial^i\psi_1\right)
             \Bigg)
\nonumber\\
&&
+P_{,XXX}X_0\Bigg(\frac{1}{3}\dot\phi_0\dot{\delta\phi}^3+X_0\alpha_1\left(-2\dot{\delta\phi}^2+2\dot\phi_0\alpha_1\dot{\delta\phi}-\frac{4}{3}X_0\alpha_1^2\right)\Bigg)
\nonumber\\
&&
+P_{,XX\phi}X_0\left(\delta\phi\dot{\delta\phi}^2-2\dot\phi_0\alpha_1\delta\phi\dot{\delta\phi}+2X_0\alpha_1^2\delta\phi\right)
+P_{,X\phi\phi}\left(\frac{1}{2}\dot\phi_0\dot{\delta\phi}-X_0\alpha_1\right){\delta\phi}^2
\nonumber\\
&&
+P_{,X}\Bigg(\alpha_1\left(-\frac{1}{2}\dot{\delta\phi}^2+\dot\phi_0\alpha_1\dot{\delta\phi}-X_0\alpha_1^2\right)
             -a^{-2}\left(\frac{1}{2}\alpha_1(\partial\delta\phi)^2+\left(\dot{\delta\phi}-\dot\phi_0\alpha_1\right)\partial_i\delta\phi\partial^i\psi_1\right)
       \Bigg)
\nonumber\\
&&+\frac{1}{2}P_{,\phi\phi}\alpha_1{\delta\phi}^2+\frac{1}{6}P_{,\phi\phi\phi}\delta\phi^3+3H^2\alpha_1^3+2a^{-2}H\alpha_1^2\partial^2\psi_1+\frac{1}{2}a^{-4}\alpha_1\left((\partial^2\psi_1)^2-\partial_i\partial_j\psi_1\partial^i\partial^j\psi_1\right)
\Bigg]. \nonumber\\
\label{3actionucurvature}
\end{eqnarray}

\subsection{Relation between gauges}
The gauges used in the previous two sections are of course related by a gauge transformation.
Introducing a new variable $\zeta_n$ defined by $\zeta_n = -H \delta \phi/\dot{\phi}_0$, $\zeta$ in the comoving gauge is related to $\delta \phi$ in the flat gauge as \cite{Maldacena:2002vr}
\begin{equation}
\zeta=\zeta_n+f(\zeta_n),
\end{equation}
where
\begin{equation} \label{redefinition}
f(\zeta)=\frac{\eta}{4c_s^2}\zeta^2+\frac{1}{c_s^2H}\zeta\dot{\zeta}+
\frac{1}{4a^2H^2}
\Big[
-(\partial\zeta)(\partial\zeta)+\partial^{-2}(\partial_i\partial_j(\partial_i\zeta\partial_j\zeta))
\Big] +
\frac{1}{2a^2H}
\Big[(\partial\zeta)(\partial\chi)-\partial^{-2}(\partial_i\partial_j
(\partial_i\zeta\partial_j\chi)) \Big] ~.
\end{equation}
On large scales where $\zeta_n$ becomes constant and we get
\begin{equation}
\zeta=\zeta_n + \frac{\eta}{4 c_s^2} \zeta_n^2.
\label{gte}
\end{equation}
We can show that this is nothing but the expression for $\zeta$ obtained in the delta-N formalism using the relations
\begin{equation}
N_{,\phi} = -\frac{H}{\dot{\phi}_0}, \quad N_{,\phi \phi} =
\frac{\dot{H}}{\dot{\phi}_0^2} - \frac{\ddot{\phi}_0 H}{\dot{\phi}_0^3},
\end{equation}
and the definition of $\eta$ in Eq.~(\ref{small}).

There are two ways to calculate the bispectrum of $\zeta$ on large scales. One is to calculate the bispectrum of $\zeta_n = - H \delta \phi/\dot{\phi}_0$ in the flat gauge and apply the delta-N formalism. It is also possible
to calculate the bispectrum of $\zeta$ in the comoving gauge. We will use both approaches in the next section.

\section{Bispectrum of curvature perturbation}
In this section, we first calculate the three point function for the field perturbations in the in-in formalism using the cubic order action obtained in the previous section.
We only consider leading order terms in slow-roll expansions. In the following, we consider two
inflation models, k-inflation and standard slow-roll inflation.
Then using the delta-N formalism, we derive the bispectrum of the curvature
perturbation. We follow the calculations in Ref.~\cite{Seery:2005gb, Arroja:2008yy, Langlois:2008wt}.
We also discuss a method to calculate it directly in the comoving gauge presented in
Ref.~\cite{Chen:2006nt}.

\subsection{Bispectrum of quantum fields in k-inflation}
First let us consider models with non-standard kinetic terms. This is known as k-inflation. In this case, the leading order terms in the slow-roll expansion in the action in the flat gauge (\ref{2actionucurvature})
and (\ref{3actiono}) are given by
\begin{equation}
S_2=\int dt d^3x
\frac{a^3 P_{,X}}{2}
\left[
\frac{1}{c_s^2}\dot{\delta \phi}^2
-\frac{1}{a^2}(\partial \delta \phi)^2
\right]
,\label{2action}
\end{equation}
\begin{equation}
S_3 =\int dt d^3x
\left[
\left(P_{,XX}\frac{\dot\phi}{2}+P_{,XXX}\frac{\dot\phi^3}{6}\right)
a^3\dot{\delta \phi}^3-P_{,XX}\frac{\dot\phi}{2}a\dot{\delta \phi}(\partial\delta \phi)^2
\right].
\label{3action}
\end{equation}
The perturbations in the interacting picture are promoted to quantum operators like
\begin{equation}
\delta \phi(\tau,\mathbf{x})=\frac{1}{(2\pi)^3}\int
d^3\mathbf{k} \delta \phi(\tau,\mathbf{k})e^{i\mathbf{k}\cdot\mathbf{x}},
\quad \delta \phi(\tau,\mathbf{k})=u(\tau,\mathbf{k})a(\mathbf{k})+u^*(\tau,-\mathbf{k})a^\dag(-\mathbf{k}).
\end{equation}
$a(\mathbf{k})$ and $a^\dag(-\mathbf{k})$ are the annihilation
and creation operator respectively, that satisfy the usual
commutation relations
\begin{equation}
\left[a(\mathbf{k_1}),a^\dag(\mathbf{k_2})\right]=(2\pi)^3\delta^{(3)}(\mathbf{k_1}-\mathbf{k_2}),
\quad
\left[a(\mathbf{k_1}),a(\mathbf{k_2})\right]=\left[a^\dag(\mathbf{k_1}),a^\dag(\mathbf{k_2})\right]=0.
\end{equation}
At leading order the solution for the mode functions is given by
\begin{equation}
u(\tau,\mathbf{k})=
\frac{H}{\sqrt{2 c_s P_{,X}}}
 \frac{1}{k^{3/2}}\left(1+ikc_s\tau\right)e^{-ikc_s\tau}.
\end{equation}

Using the in-in formalism Eq.~(\ref{in-in}),
the vacuum expectation value of the three point operator in the
interaction picture is written as
\cite{Maldacena:2002vr,Weinberg:2005vy}
\begin{equation}
\langle \delta \phi(t,\mathbf{k_1}) \delta \phi(t,\mathbf{k_2})
\delta \phi(t,\mathbf{k_3}) \rangle
=-i\int^t_{t_0}d\tilde
t \langle
\left[\delta \phi(t,\mathbf{k_1}) \delta \phi(t,\mathbf{k_2})
\delta \phi(t,\mathbf{k_3}),H_I(\tilde
t)\right]\rangle,
\end{equation}
where $t_0$ is some early time during inflation when the field's
vacuum fluctuation are deep inside the horizons, $t$ is some time
after horizon exit. If one
uses conformal time, it's a good approximation to perform the
integration from $-\infty$ to $0$ because $\tau\approx-(aH)^{-1}$.
$H_I$ denotes the interaction Hamiltonian and it is given by
$H_I=-L_3$, where $L_3$ is the Lagrangian obtained from the action
(\ref{3action}). using the solution for the mode function
and commutation relations for the creation and annihilation operators,
we get
\begin{equation}
\langle \delta \phi(\mathbf{k_1}) \delta \phi(\mathbf{k_2})
\delta \phi(\mathbf{k_3})
\rangle=
-(2\pi)^3\delta^{(3)}(\mathbf{k_1}+\mathbf{k_2}+\mathbf{k_3})
\frac{H^4}{\sqrt{2 \epsilon} c_s^2 (P_{,X})^{3/2}}
\frac{1}{\Pi_{i=1}^3k_i^3} {\cal A}^{k-inf}_{\phi}(k_1, k_2, k_3),
\end{equation}
where
\begin{eqnarray}
{\cal A}^{k-inf}_{\phi} &=&
-\frac{3 \lambda}{\Sigma} \frac{k_1^2k_2^2k_3^2}{K^3}
+ \left(\frac{1}{c_s^2}-1 \right)
\frac{k_1^2\mathbf{k_2}\cdot\mathbf{k_3}}{K}
\left(1+\frac{k_2+k_3}{K}+2\frac{k_2k_3}{K^2}\right)
+2\,\,\mathrm{cyclic\, terms} \\
&=&
\left(\frac{1}{c_s^2}-1 - \frac{2 \lambda}{\Sigma} \right)
\frac{3 k_1^2 k_2^2 k_3^2}{K}
+\frac{1-c_s^2}{c_s^2}
\left(
-\frac{1}{K} \sum_{i>j} k_i^2 k_j^2 + \frac{1}{2 K^2}
\sum_{i \neq j} k_i^2 k_j^3 + \frac{1}{8} \sum_i k_i^3
\right).
\label{kinf}
\end{eqnarray}

\subsection{Bispectrum of quantum fields in slow-roll inflation}
In slow-roll inflation with a standard kinetic term $P(X)= X- V(\phi)$,
leading-order terms in the third-order action (\ref{3actiono})
are given by
\begin{eqnarray}
\label{S3}
  \nonumber
  S_3 = \int dt \, dx^3 \; a^3
  \Bigg( - \frac{1}{a^2}  \dot{\delta \phi} \partial\psi
  \partial \delta \phi- \frac{1}{4H}  \dot{\phi} \delta \phi  (\dot{\delta \phi})^2
  - \frac{1}{a^4} \frac{1}{4H}  \dot{\phi} \delta \phi
  (\partial \delta \phi)^2 \Bigg) ,
\end{eqnarray}
where $\psi$ was defined in Eq.~(\ref{N1orderphi}) and is given to leading order by
\begin{equation}
  \label{threept:psi}
  \partial^2 \psi = - \frac{a^2}{2H}  \dot{\phi} \dot{\delta \phi}.
\end{equation}
The three point function can be calculated in the same way. We obtain
\begin{equation}
\langle \delta \phi(\mathbf{k_1}) \delta \phi(\mathbf{k_2}) \delta \phi(\mathbf{k_3})\rangle=-
(2\pi)^3\delta^{(3)}(\mathbf{k_1}+\mathbf{k_2}+\mathbf{k_3})\frac{H^4}{\sqrt{2 \epsilon}}
\frac{1}{\Pi_{i=1}^3k_i^3} {\cal A}_{\phi}^{stand}(k_1, k_2, k_3),
\end{equation}
where
\begin{equation}
A_{\phi}^{stand} (k_1, k_2, k_3) =
\epsilon\left( -\frac{1}{8}
\sum_i k_i^3 + \frac{1}{8}
\sum_{i\neq j} k_i k_j^2 + \frac{1}{K} \sum_{i>j} k_i^2 k_j^2
\right)
.
\label{Aslow}
\end{equation}

\subsection{Bispectrum of curvature perturbation}
Now we can apply the delta-N formalism to calculate the bispectrum of the curvature
perturbation. We define the bispectrum of the curvature perturbation as
\begin{eqnarray}
\langle \zeta(\textbf{k}_1)\zeta(\textbf{k}_2)\zeta(\textbf{k}_3)\rangle
&=&
(2\pi)^7\delta^3(\textbf{k}_1+\textbf{k}_2+\textbf{k}_3)
({\cal P}_{\zeta})^2
\frac{1}{\prod_i k_i^3}~ \CA_{\zeta}(k_1,k_2,k_3) ~,
\label{3pointz}
\end{eqnarray}
where ${\cal P}_{\zeta}$ is given by Eq.~(\ref{power1}).
In k-inflation models,
in the small sound speed limit and at the leading order in slow-roll expansion,
the relation between the curvature perturbation and the field perturbation is
simply given by $\zeta = - H \delta \phi/\dot{\phi}  =
P_{,X} \delta \phi /2 \epsilon$.
Then the three point function for $\zeta$ is given by Eq.~(\ref{3pointz})
where ${\cal A}_{\zeta} ={\cal A}_{\phi}^{k-inf}$
given by Eq.~(\ref{kinf}).

In standard slow-roll inflation case, the relation between $\zeta$ and $\zeta_n$ can
be written as $\zeta =\zeta_n + \eta \zeta_n^2/4$. Then the the bispectrum
of the curvature perturbation is given by Eq.~(\ref{3pointz})
where $\CA_{\zeta}$ is given by
\begin{equation}
\CA_{\zeta} =
\CA^{stand}_{\phi} +
\eta \left( \frac{1}{8} \sum_i k_i^3
\right).
\label{zetaslow}
\end{equation}

\subsection{Computation in comoving gauge}
It is also possible to calculate the bispectrum of the curvature perturbation in
comoving gauge and this gives a very useful consistency check.
Here we follow Ref.~\cite{Chen:2006nt} and see how this works.

In fact the cubic effective action in (\ref{3action}) looks like order
$\CO(\epsilon^0)$ in the slow variation parameters while in the previous
section, we find that the bispectrum is suppressed by slow-roll parameters
in slow-roll inflation. In slow-roll inflation, as
emphasized and demonstrated in Ref.~\cite{Maldacena:2002vr},
one can perform a
lot of integrations by parts and cancel terms of order
$\CO(\epsilon^0)$ and $\CO(\epsilon)$. The resulting cubic action is
actually of leading order $\CO(\epsilon^2)$ in slow roll parameters.
A similar analysis can be performed for the general
Lagrangian in Ref.~\cite{Seery:2005wm}.
Except for terms that
are proportional to $1-c_s^2$ or $\lambda$, the rest of the terms can
be cancelled to the second order $\CO(\epsilon^2)$ and the cubic order action Eq.~(\ref{3actionz})
can be rewritten as
\begin{eqnarray} \label{action3}
S_3&=&\int dt d^3x \Big[
-a^3 \Big(\Sigma(1-\frac{1}{c_s^2})+2\lambda \Big)\frac{\dot{\zeta}^3}{H^3}
+\frac{a^3\epsilon}{c_s^4}(\epsilon-3+3c_s^2)\zeta\dot{\zeta}^2
+
\frac{a\epsilon}{c_s^2}(\epsilon-2s+1-c_s^2)\zeta(\partial\zeta)^2
\nonumber \\
&& -
2a \frac{\epsilon}{c_s^2}\dot{\zeta}(\partial
\zeta)(\partial \chi) +
\frac{a^3\epsilon}{2c_s^2}\frac{d}{dt}
\left(\frac{\eta}{c_s^2} \right)
\zeta^2\dot{\zeta}
+\frac{\epsilon}{2a}(\partial\zeta)(\partial
\chi) \partial^2 \chi +\frac{\epsilon}{4a}(\partial^2\zeta)(\partial
\chi)^2+ 2 f(\zeta) \left.\frac{\delta L}{\delta \zeta} \right|_1 \Big] ~,
\end{eqnarray}
where $\chi$ is defined in Eq.~(\ref{N1order}) and in the last term
\begin{eqnarray}
\left. \frac{\delta
L}{\delta\zeta} \right |_1 &=& a
\left( \frac{d\partial^2\chi}{dt}+H\partial^2\chi
-\epsilon\partial^2\zeta \right).
\end{eqnarray}
Here $\partial^{-2}$ is the inverse Laplacian,
$\delta
L/\delta\zeta|_1$ is the variation of the
quadratic action with respect to the perturbation $\zeta$, therefore
the last term which is proportional to $\delta
L/\delta\zeta|_1$ can be absorbed by a field redefinition of
$\zeta$. It can be easily shown that the field redefinition that
absorbs this term is
\begin{eqnarray}
\zeta \rightarrow \zeta_n+f(\zeta_n), ~
\end{eqnarray}
where $f(\zeta)$ is given by Eq.~(\ref{redefinition}).
This is nothing but the relation between $\zeta$ and $\zeta_n$ obtained from
the guage transformation between the flat gauge and comoving guage.
One then computes the vacuum expectation value of the three point
function in the interaction picture in the same way. We get Eq.~(\ref{3pointz})
with
\begin{eqnarray}
\CA &=& \left(\frac{1}{c_s^2}-1
-\frac{2\lambda}{\Sigma} \right) \frac{3k_1^2k_2^2k_3^2}{2K^3}
+
\left(\frac{1}{c_s^2}-1\right)
\left(-\frac{1}{K}\sum_{i>j}k_i^2k_j^2+\frac{1}{2K^2}
\sum_{i\neq j}k_i^2k_j^3+\frac{1}{8}\sum_{i}k_i^3 \right)
\nonumber \\
&+& \frac{\epsilon}{c_s^2} \left( -\frac{1}{8}
\sum_i k_i^3 + \frac{1}{8}
\sum_{i\neq j} k_i k_j^2 + \frac{1}{K} \sum_{i>j} k_i^2 k_j^2
\right)
+ \frac{\eta}{c_s^2} \left( \frac{1}{8} \sum_i k_i^3
\right)
\nonumber \\
&+& \frac{s}{c_s^2} \left( -\frac{1}{4} \sum_i k_i^3 - \frac{1}{K}
\sum_{i>j} k_i^2 k_j^2 +
\frac{1}{2K^2} \sum_{i\neq j} k_i^2 k_j^3 \right) ~.
\label{3pointPre}
\end{eqnarray}
In k-inflation, the first two terms are the leading order contributions in slow-roll expansions. The remaining terms are ${\cal O}(\epsilon)$. Note that one should take
into account ${\cal O}(\epsilon)$ corrections from the leading order contribution in order to
obtain a full expression up to ${\cal O}(\epsilon)$ in k-inflation. In standard slow-roll inflation, the first two terms and the last term vanishes and $c_s=1$.

\section{Shapes of Bispectrum}
In this section, we compare the prediction of the bispectrum in k-inflation with the local-type non-Gaussianity.
The discussions in this section are based on Refs.~\cite{Babich:2004gb, Creminelli:2005hu}.

The bispectrum in the local-type non-Gaussianity is often characterized by
\begin{equation}
\zeta = \zeta_n + \frac{3}{5} f_{NL}^{local} (\zeta_n^2 - \langle \zeta_n^2 \rangle),
\label{fnldef}
\end{equation}
where $\zeta_n$ obeys Gaussian statistics.
Originally the $f_{NL}^{local}$ parameter was introduce to parametrise a non-linearity in the curvature perturbation $\Phi$ in the Longitudinal gauge which is related to $\zeta$ as
$\Phi = (3/5) \zeta$. Note that Ref.~\cite{Maldacena:2002vr} uses a different sign convention for $f_{NL}^{local}$ from
WMAP papers (see \cite{Komatsu:2010fb} for the latest result).
Here we follow the definition used in WMAP papers. Eq.~(\ref{fnldef}) is nothing more than
the expression for $\zeta$ in the delta-N formalism. In slow-roll inflation $f_{NL}^{local}$ is
${\cal O}(\epsilon)$ but models like curvaton predicts $f_{NL}^{local}$ larger than one. The bispectrum of curvature perturbation is given by
\begin{equation}
\langle \zeta(\textbf{k}_1)\zeta(\textbf{k}_2)\zeta(\textbf{k}_3)\rangle
=
(2\pi)^3\delta^3(\textbf{k}_1+\textbf{k}_2+\textbf{k}_3)
({\cal P}_{\zeta})^2
~ F(k_1, k_2, k_3),
\end{equation}
where
\begin{equation}
F_{local}(k_1,k_2,k_3)= (2 \pi)^4 \left(
\frac{3}{10} f_{NL}^{local}  \right)
\left(\frac{1}{k_1^3 k_2^3} + \frac{1}{k_2^3 k_3^3}
+ \frac{1}{k_3^3 k_1^3}   \right).
\label{local}
\end{equation}

Eq.~(\ref{fnldef}) describes (at leading order) the most generic form
of non-Gaussianity which is local in real space. This form is therefore expected
for models where non-linearities develop outside the horizon.  This
happens for all the models in which the fluctuations of an additional
light field, different from the inflaton, contribute to the curvature
perturbations we observe. In this case non-linearities come from the
evolution of this field outside the horizon and from the conversion
mechanism which transforms the fluctuations of this field into the curvature
perturbations. Both these sources of non-linearity give a
non-Gaussianity of the form (\ref{fnldef}) because they occur outside
the horizon. Examples of this general scenario are the curvaton models
\cite{Lyth:2002my}, models with fluctuations in the reheating
efficiency \cite{Dvali:2003em,Dvali:2003ar} and multi-field
inflationary models \cite{Bernardeau:2002jy}.

Being local in position space, Eq.~(\ref{fnldef}) describes correlation among
Fourier modes of very different
$k$. It is instructive to take the limit in which one of the modes becomes of very long
wavelength \cite{Maldacena:2002vr}, $k_3 \rightarrow 0$, which implies, due to momentum
conservation, that the other
two $k$'s become equal and opposite.
The long wavelength mode $\zeta_{\vec k_3}$ freezes out much before the others and behaves as a
background for their evolution.
In this limit $F_{local}$ is proportional to the power spectrum of the short and long
wavelength modes
\begin{equation}
\label{eq:limitloc}
F_{local} \propto \frac1{k_3^3} \frac1{k_1^3} \;.
\end{equation}
This means that the short wavelength 2-point function $\langle \zeta_{\vec k_1} \zeta_{-\vec k_1} \rangle$ depends
linearly on the long wavelength mode $\zeta_{\vec k_3}$
\begin{equation}
\label{eq:linear}
\langle \zeta_{\vec k_3}\zeta_{\vec k_1}\zeta_{-\vec k_1} \rangle \propto
\langle \zeta_{\vec k_3}\zeta_{-\vec k_3}\rangle \frac\partial{\partial \zeta_{\vec k_3}} \langle
\zeta_{\vec k_1}\zeta_{-\vec k_1} \rangle \;.
\end{equation}
From this point of view we expect that any bispectra will reduce to the local shape (\ref{local}) in the
degenerate limit we considered
if the derivative with respect to the long wavelength mode does not vanish.

In standard single field slow-roll inflation, as pointed
out in \cite{Maldacena:2002vr}, different points along the background wave are equivalent to shift in time
along the inflaton trajectory, so that the derivative with respect to the background wave is proportional to the
tilt of the scalar spectrum. This can be explicitly checked in the full expression of the 3-point function
(Eqs.~(\ref{Aslow}) and (\ref{zetaslow})):
\begin{equation}
\label{eq:malda}
F_{stand}(k_1, k_2,k_3) = \frac{(2\pi)^4}{\prod_i k_i^3}
\left[
\epsilon \left( -\frac{1}{8}
\sum_i k_i^3 + \frac{1}{8}
\sum_{i\neq j} k_i k_j^2 + \frac{1}{K} \sum_{i>j} k_i^2 k_j^2
\right)
+ \eta \left( \frac{1}{8} \sum_i k_i^3
\right)\right] \;,
\end{equation}
In the limit $k_3 \rightarrow 0$ Eq.~(\ref{eq:malda}) goes as
\begin{equation}
\label{eq:maldalimit}
F_{stand}(k_3 \rightarrow 0) \propto (2\epsilon + \eta) \frac1{k_3^3 k_1^3} =
-(n_s-1)\frac1{k_3^3 k_1^3} \;.
\end{equation}
As expected the tilt in the spectrum $n_s$ fixes the degenerate limit of the 3-point function. Note however
that expression (\ref{eq:malda}) is not of the local form (\ref{local}) but contains contributions
which are important for non-degenerate triangles. If we compare expression (\ref{local}) and (\ref{eq:malda})
and neglect the different shape dependence, we see that standard single-field inflation predicts $f_{NL}^{local}$
of order of the slow-roll parameters.

We have seen that the degenerate limit $k_3 \rightarrow 0$ describes the
effect of a slowly-varying long-wavelength perturbation on the 2-point function
of short wavelength modes.
In many models, the correlation is much weaker in this limit than in the
local model (\ref{local}). Physically this means that the correlation
is among modes with comparable wavelength which go out of the horizon
nearly at the same time. In this case the 3-point function in the
degenerate limit is suppressed by powers of $k_3$ with respect to the
behaviour of Eq.~(\ref{eq:limitloc}). We have correlation among modes
of comparable wavelength in all models in which the non-Gaussianity is
generated by derivative interactions: these interactions become
exponentially irrelevant when the modes go out of the horizon because
both time and spatial derivatives become small, so that all the
correlation is among modes freezing almost at the same time.

\begin{figure}[t!]
\begin{center}
\includegraphics[width=7cm]{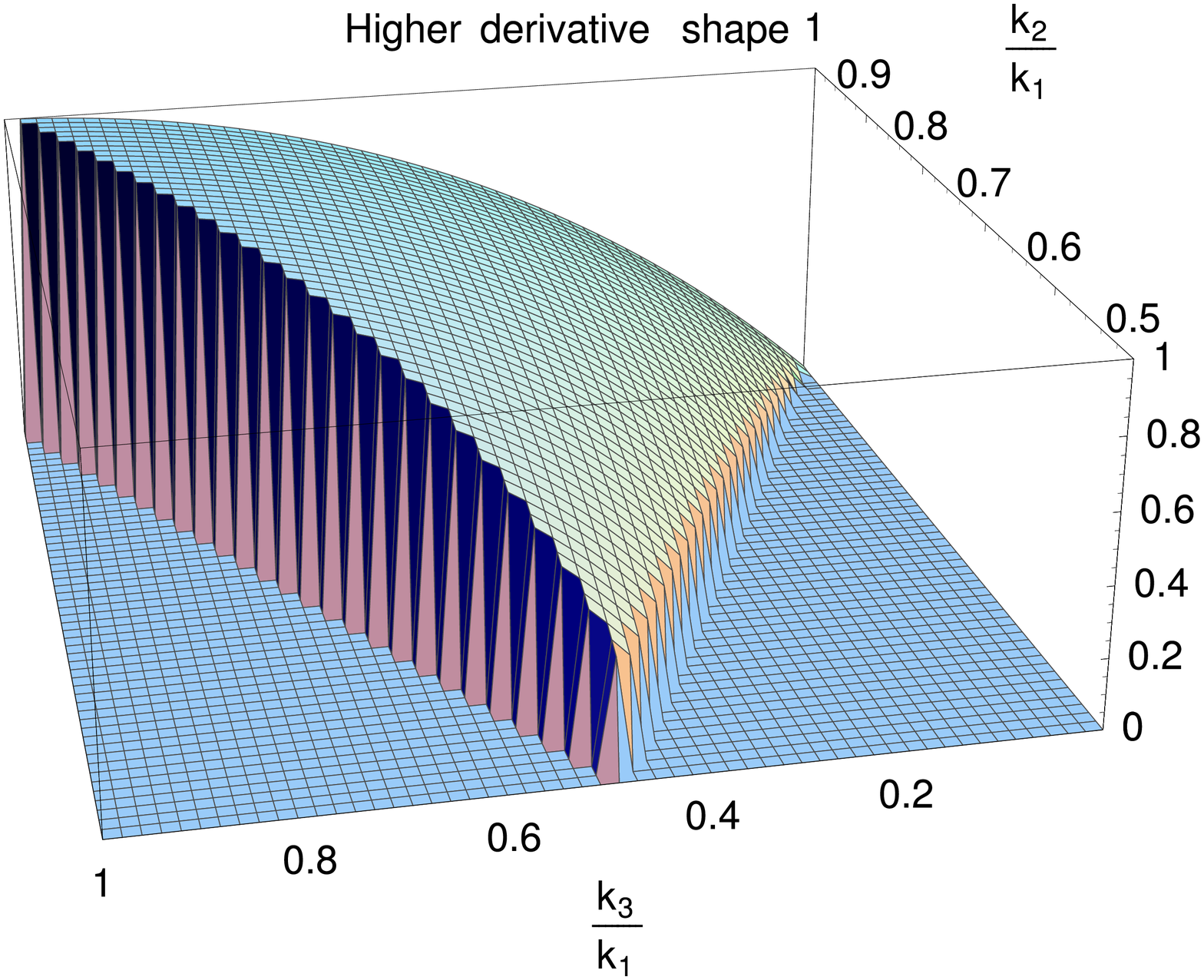}
\includegraphics[width=7cm]{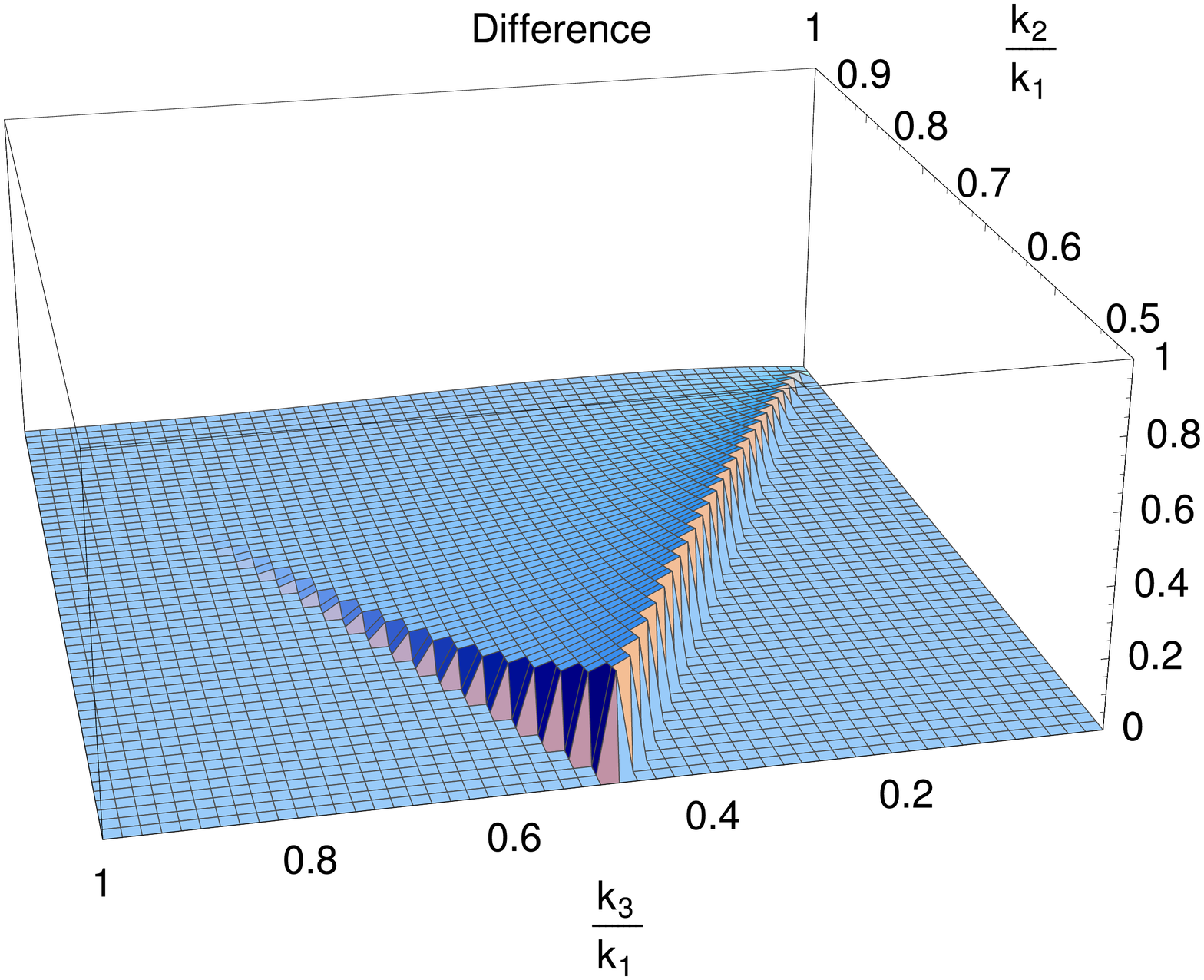}
\caption{\label{fig:hddiff} \small  {\it Left:} Plot of the function $F(1,\, k_2/k_1, \, k_3/k_1) (k_2/k_1)^2 (k_3/k_1)^2$
predicted by the DBI models \cite{Alishahiha:2004eh}.
{\it Right:} Difference between the above plot and the analogous one (top of fig.~\ref{fig:shapes}) for the factorizable
equilateral shape used in the analysis. From \cite{Creminelli:2005hu}.}
\end{center}
\end{figure}

K-inflation is a typical example for these type of models. The three point function is
obtained in Eq.~(\ref{3pointPre}) and given by
\begin{equation}
F(k_1,k_2,k_3) = \frac{(2\pi)^4}{\prod_i k_i^3}\left[
\left(\frac{1}{c_s^2}-1
-\frac{2\lambda}{\Sigma} \right) \frac{3k_1^2k_2^2k_3^2}{2K^3}
+
\left(\frac{1}{c_s^2}-1\right)
\left(-\frac{1}{K}\sum_{i>j}k_i^2k_j^2+\frac{1}{2K^2}
\sum_{i\neq j}k_i^2k_j^3+\frac{1}{8}\sum_{i}k_i^3 \right)
\right].
\label{DBI}
\end{equation}
In a model of inflation based on the DBI action,
\begin{equation}
P(\phi, X) = -f(\phi)^{-1} \sqrt{1- 2 X f(\phi)} +f(\phi)^{-1}+ V(\phi),
\end{equation}
$\lambda$ is given by
\begin{equation}
\lambda
=\frac{\Sigma}{2} \left(\frac{1}{c_s^2}-1 \right).
\end{equation}
Thus the first term vanishes in Eq.~(\ref{DBI}).
Unfortunately, a function $F$ is not factorizable, so it is not easy to perform an optimal
analysis using CMB observations.
However, it is a very good approximation to take a
factorizable shape function $F$ which is close to Eq.~(\ref{DBI})
and perform the analysis for this shape. In the limit $k_1 \to 0$ with $k_2$ and $k_3$ fixed, all the equilateral
functions diverge as  $k_1^{-1}$ \cite{Babich:2004gb} (while the local form eq.~(\ref{local})
goes as $k_1^{-3}$). The factorizable function that satisfies this condition is given by
\begin{equation}
\label{eq:ours}
F(k_1,k_2,k_3) = (2 \pi)^4 \left(\frac{9 }{10}f_{NL}^{equil}\right)
 \left(-\frac1{k_1^3 k_2^3} -
\frac1{k_1^3 k_3^3} - \frac1{k_2^3 k_3^3}  - \frac2{k_1^2 k_2^2 k_3^2} + \frac1{k_1 k_2^2 k_3^3}
+ (5 \; perm.) \right) \;,
\end{equation}
where the permutations act only on the last term in parentheses.
In figure \ref{fig:hddiff} we study the equilateral function predicted in DBI inflation
(\ref{DBI}). In the second part of the figure we show the difference between this function and the factorizable one used in our analysis.
We see that the relative difference is quite small. The same remains true for other equilateral shapes
(see \cite{Babich:2004gb} for the analogous plots for other models).

\begin{figure}[th!!]
\begin{center}
\includegraphics[width=7cm]{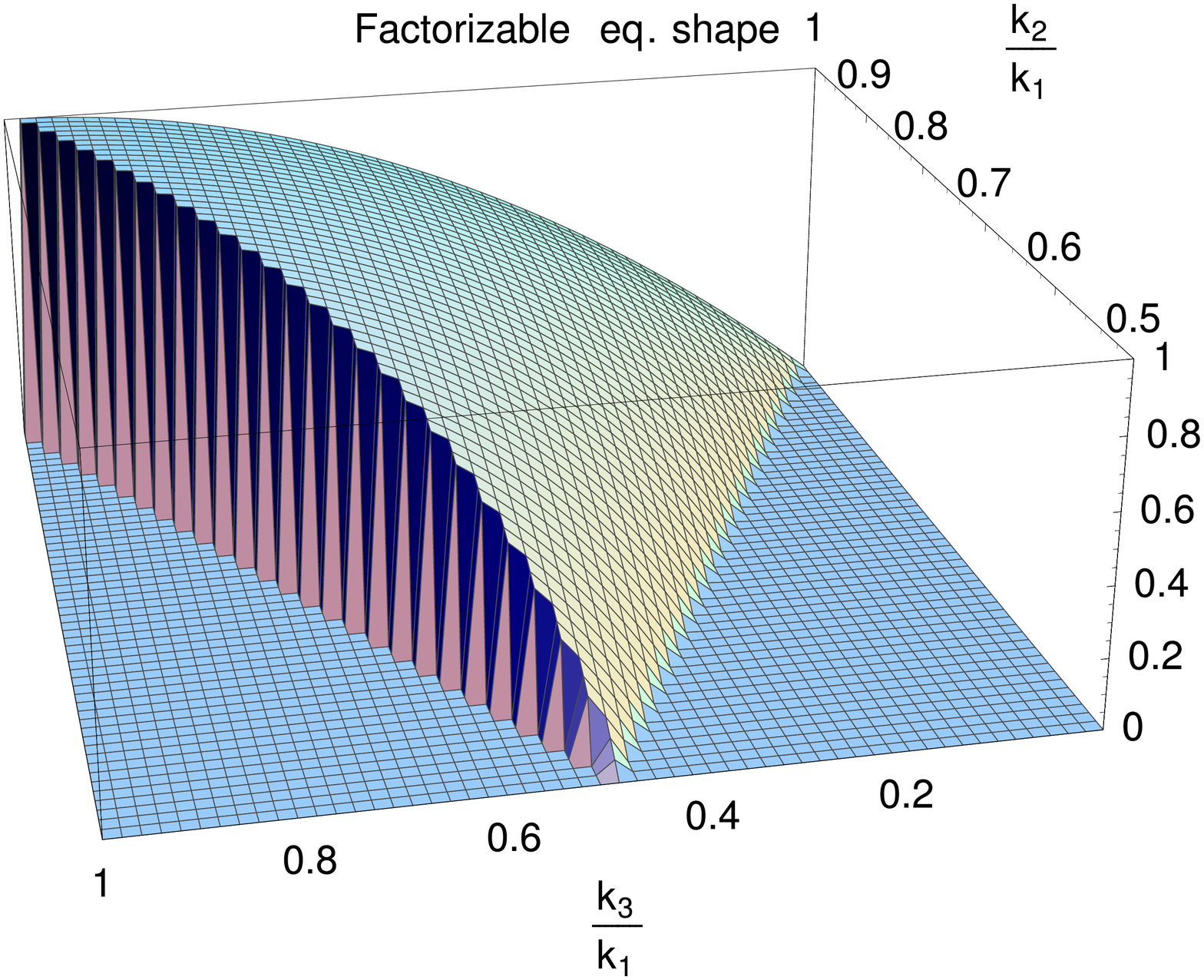}
\includegraphics[width=7cm]{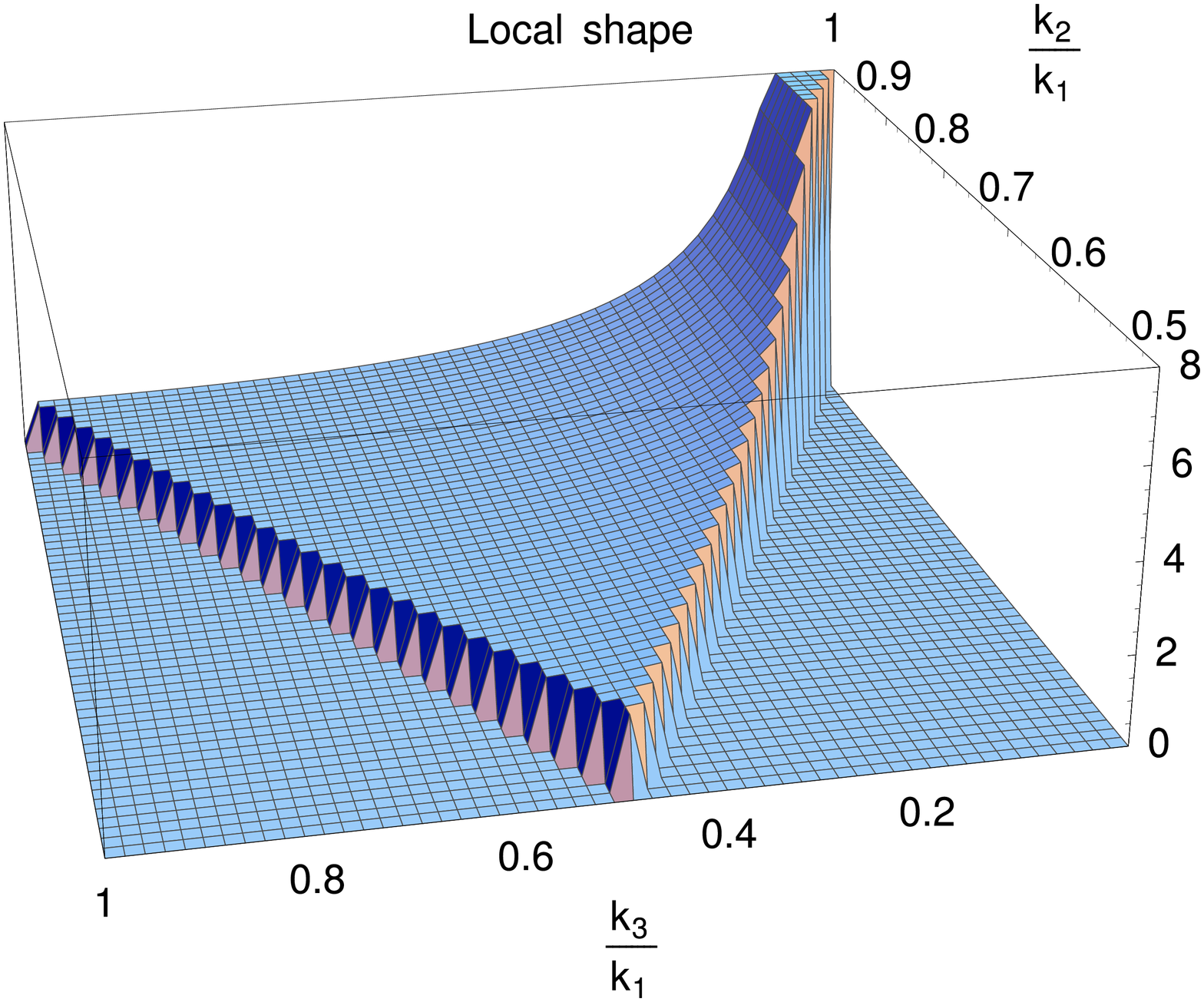}
\caption{\label{fig:shapes} \small  Plot of the function
$F(1,\, k_2/k_1, \, k_3/k_1) (k_2/k_1)^2 (k_3/k_1)^2$ for the
equilateral shape used in the analysis (left) and for the local shape (right). The functions are both
normalized to unity for equilateral configurations $\frac{k_2}{k_1}= \frac{k_3}{k_1}=1$.
Since $F(k_1,k_2,k_3)$ is symmetric in its three arguments,
it is sufficient to specify it for $k_1\ge k_2\ge k_3$, so $\frac{k_3}{k_1} \le \frac{k_2}{k_1} \le 1$ above.
Moreover, the triangle inequality says that no side can be longer than the sum of the other two,
so we only plot $F$ in the triangular region $1-\frac{k_2}{k_1}\leq \frac{k_3}{k_1} \leq\frac{k_2}{k_1} \le 1$ above,
setting it to zero elsewhere. From \cite{Creminelli:2005hu}.}
\end{center}
\end{figure}

In figure \ref{fig:shapes}, we compare this function with the local shape. The dependence of both functions
under a common rescaling of all $k$'s is
fixed to be $\propto k^{-6}$ by scale invariance, so that we can factor out $k_1^{-6}$ for example.
Everything will now depend only on the ratios $k_2/k_1$ and $k_3/k_1$, which fix the shape of the
triangle in momentum space. For each shape we plot $F(1,k_2/k_1,k_3/k_1) (k_2/k_1)^2 (k_3/k_1)^2$;
this is the relevant quantity if we are interested in the relative importance of different triangular
shapes. The square of this function gives the signal to noise contribution of a particular
shape in momentum space \cite{Babich:2004gb}. We see that for the function (\ref{eq:ours}), the signal
to noise is concentrated on equilateral configurations, while squeezed triangles with one side much
smaller than the others are the most relevant for the local shape.

\section{Conclusion}
In this review, the bispectrum of curvature perturbation is calculated using the in-in formalsim and the delta-N formalism. There are two distinct contributions to the bispectrum. One is coming from a non-linear relation between the curvature perturbation $\zeta$ and quantum fluctuations of a field at the horizon crossing. In this case the non-linearities come from the evolution of this field outside the horizon. Being local in a position space, the shape of the bispectrum is highly non-local in a Fourie space having a maximum signal for the squeezed configuration $k_3 \ll k_1, k_2$. The other contribution is coming from the bispectrum of quantum fields generated under horizon scales. In models in which the non-Gaussianity is
generated by derivative interactions such as DBI inflation and k-inflation models,
we have correlation among modes of comparable wavelength in all models and these
interactions become exponentially irrelevant when the modes go out of the horizon because
both time and spatial derivatives become small, so that all the
correlation is among modes freezing almost at the same time. Then the bispectrum has a peak at the equilateral configuration $k_1 \sim k_2 \sim k_3$.

In order to put constraints on the bispectrum from CMB observations, it is necessary to construct an estimator that uses a model prediction for the bispectrum as an template. For the local-type non-Gaussianity and the equilateral non-Gaussianity, the constraints obtained in WMAP 7-year results are \cite{Komatsu:2010fb}
\begin{equation}
-10 <f_{NL}^{local} < 74, \quad -214 < f_{NL}^{equil} <266.
\end{equation}
at $95\%$ confidence level.
Recently, it has been found that Large Scale Structure (LSS) can give a similar level of constraints on the local type non-Gaussianity from the scale-dependent bias effects on the halo power spectrum
\cite{Dalal:2007cu}
while this effect is absent in the equilateral non-Gaussianity \cite{Taruya:2008pg, Verde:2009hy}.
The constraint on $f_{NL}^{local}$ is obtained from SDSS as $-29 < f_{NL}^{local} <70$ \cite{Slosar:2008hx} and combining it to WMAP 7-year results, we get
$-5 <f_{NL}^{local} < 59$ \cite{Komatsu:2010fb}.
Thus currents observations are consistent with Gaussian primordial curvature perturbations but future experiments such as Planck will give much tighter constraints
$f_{NL}^{local} \sim {\cal O}(1)$ and we may be able to detect a deviation from Gaussianity which has a huge impact on early universe models.

There are a lot issues that are not covered by this review. We will mention some of the issues here:

\begin{itemize}

\item {\it Trispectrum}\\
In this review, we concentrated on the leading order non-Gaussianity, i.e. the bispectrum but it has been recognized that the trispectrum could give a useful information to distinguish between many possible models that predict large non-Gaussianity. For the local non-Gaussianity, we can easily extend the model by expanding $\zeta$ up to the third order $\zeta = \zeta_n + f_{NL} \zeta_n^2/2 + g_{NL} \zeta_n^3/6$. The trispetrum is characterized by two parameters $\tau_{NL}=f_{NL}^2$ and $g_{NL}$
\cite{Seery:2006js, Seery:2006vu, Byrnes:2006vq}. The constraints on these parameters are rather weak, $- 631 < \sqrt{f_{NL}^2} < 717$ and $-3.80 <g_{NL}/10^6 <3.88$ from WMAP 5-year results at $95\%$ confidence level
\cite{Smidt:2010sv}. But again the Planck will improve these significantly \cite{Kogo:2006kh}.
The full trispectrum in DBI inflation at the leading order in small sound speed limit has been obtained \cite{Huang:2006eh, Chen:2009bc, Arroja:2009pd, RenauxPetel:2009sj}. Unlike the bispectrum, there are still two degrees of freedom even for the equilateral configurations $k_1 \sim k_2 \sim k_3 \sim k_4$ and also the form of the trispectrum is too complicated to be used for the estimator. It is necessary to develop approximations for the shape of the trispectrum in DBI inflation as is done for the bispectrum.

\item{\it Multi-field inflation}\\
In single field inflation models with a standard kinetic term, the resulting non-Gaussianity is small suppressed by slow-roll parameters. However, in multi-field models, it is possible to have large local non-Gaussianity due to non-linear dynamics of fields outside horizon. The delta-N formalism is easy to be extended to multi-field models and there have been extensive study of non-Gaussianity in multi-field models using the delta-N formalism (see for example \cite{Byrnes:2009qy} and references therein). Multi field effects are also important in DBI inflation. In DBI inflation, fluctuations along the entropy directions of the fields that are orthogonal to the field trajectory have the same sound speed as the adiabatic fluctuations along the field trajectory \cite{Langlois:2008wt, Langlois:2008qf, Arroja:2008yy, Mizuno:2009cv}. If the trajectory makes a turn in a field space, this converts the entropy perturbations to the curvature perturbations. Although the bispectrum is enhanced by this conversion, the enhancement of the power spectrum is stronger and $f_{NL}^{equil}$ becomes smaller in multi-field models, which help ease stringent constraints on DBI inflation models in string theory \cite{Langlois:2008wt}.
It has been shown that the trispectrum is enhanced for a given $f_{NL}^{equil}$ \cite{Mizuno:2009mv, Gao:2009at}.
It is also possible that the multi-field effects modify the bipsectrum for quantum field at the horizon crossing. In the so-called quasi-single inflation models \cite{Chen:2009zp},
the entropy perturbations develops large non-Gaussianity. The conversion of entropy perturbations to the curvature perturbation can happen near the horizon crossing and during this transition the shape of the bispectrum can be modified in a non-trivial way.
\end{itemize}

There are many other possibilities to get large non-Gaussianity of quantum fields such as a feature in inflaton potentials \cite{Chen:2006xjb, Chen:2008wn}. All these models predict distinct shapes of the bispectrum and trispectrum. In the future, we may be able to exploit CMB and LSS data to distinguish between many possible early universe models via non-Gaussianity.

\begin{acknowledgements}
We would like to thank the authors of Ref.~\cite{Creminelli:2005hu} for giving us a permission 
to use Fig.1 and Fig.2 in this article. KK is supported by the UK's Science \& Technology
Facilities Council, the European Research Council and Research Councils UK.
\end{acknowledgements}


\begin{thebibliography}{97}
\expandafter\ifx\csname natexlab\endcsname\relax\def\natexlab#1{#1}\fi
\expandafter\ifx\csname bibnamefont\endcsname\relax
  \def\bibnamefont#1{#1}\fi
\expandafter\ifx\csname bibfnamefont\endcsname\relax
  \def\bibfnamefont#1{#1}\fi
\expandafter\ifx\csname citenamefont\endcsname\relax
  \def\citenamefont#1{#1}\fi
\expandafter\ifx\csname url\endcsname\relax
  \def\url#1{\texttt{#1}}\fi
\expandafter\ifx\csname urlprefix\endcsname\relax\def\urlprefix{URL }\fi
\providecommand{\bibinfo}[2]{#2}
\providecommand{\eprint}[2][]{\url{#2}}

\bibitem[{\citenamefont{http://www.rssd.esa.int/index.php?project=Planck}(2010%
)}]{PLANCK}
\bibinfo{author}{\bibnamefont{http://www.rssd.esa.int/index.php?project=Planck%
}} (\bibinfo{year}{2010}).

\bibitem[{\citenamefont{Acquaviva et~al.}(2003)\citenamefont{Acquaviva,
  Bartolo, Matarrese, and Riotto}}]{Acquaviva:2002ud}
\bibinfo{author}{\bibfnamefont{V.}~\bibnamefont{Acquaviva}},
  \bibinfo{author}{\bibfnamefont{N.}~\bibnamefont{Bartolo}},
  \bibinfo{author}{\bibfnamefont{S.}~\bibnamefont{Matarrese}},
  \bibnamefont{and} \bibinfo{author}{\bibfnamefont{A.}~\bibnamefont{Riotto}},
  \bibinfo{journal}{Nucl. Phys.} \textbf{\bibinfo{volume}{B667}},
  \bibinfo{pages}{119} (\bibinfo{year}{2003}), \eprint{astro-ph/0209156}.

\bibitem[{\citenamefont{Maldacena}(2003)}]{Maldacena:2002vr}
\bibinfo{author}{\bibfnamefont{J.~M.} \bibnamefont{Maldacena}},
  \bibinfo{journal}{JHEP} \textbf{\bibinfo{volume}{05}}, \bibinfo{pages}{013}
  (\bibinfo{year}{2003}), \eprint{astro-ph/0210603}.

\bibitem[{\citenamefont{Linde and Mukhanov}(1997)}]{Linde:1996gt}
\bibinfo{author}{\bibfnamefont{A.~D.} \bibnamefont{Linde}} \bibnamefont{and}
  \bibinfo{author}{\bibfnamefont{V.~F.} \bibnamefont{Mukhanov}},
  \bibinfo{journal}{Phys. Rev.} \textbf{\bibinfo{volume}{D56}},
  \bibinfo{pages}{535} (\bibinfo{year}{1997}), \eprint{astro-ph/9610219}.

\bibitem[{\citenamefont{Bartolo et~al.}(2002)\citenamefont{Bartolo, Matarrese,
  and Riotto}}]{Bartolo:2001cw}
\bibinfo{author}{\bibfnamefont{N.}~\bibnamefont{Bartolo}},
  \bibinfo{author}{\bibfnamefont{S.}~\bibnamefont{Matarrese}},
  \bibnamefont{and} \bibinfo{author}{\bibfnamefont{A.}~\bibnamefont{Riotto}},
  \bibinfo{journal}{Phys. Rev.} \textbf{\bibinfo{volume}{D65}},
  \bibinfo{pages}{103505} (\bibinfo{year}{2002}), \eprint{hep-ph/0112261}.

\bibitem[{\citenamefont{Bernardeau and Uzan}(2002)}]{Bernardeau:2002jy}
\bibinfo{author}{\bibfnamefont{F.}~\bibnamefont{Bernardeau}} \bibnamefont{and}
  \bibinfo{author}{\bibfnamefont{J.-P.} \bibnamefont{Uzan}},
  \bibinfo{journal}{Phys. Rev.} \textbf{\bibinfo{volume}{D66}},
  \bibinfo{pages}{103506} (\bibinfo{year}{2002}), \eprint{hep-ph/0207295}.

\bibitem[{\citenamefont{Bernardeau and Uzan}(2003)}]{Bernardeau:2002jf}
\bibinfo{author}{\bibfnamefont{F.}~\bibnamefont{Bernardeau}} \bibnamefont{and}
  \bibinfo{author}{\bibfnamefont{J.-P.} \bibnamefont{Uzan}},
  \bibinfo{journal}{Phys. Rev.} \textbf{\bibinfo{volume}{D67}},
  \bibinfo{pages}{121301} (\bibinfo{year}{2003}), \eprint{astro-ph/0209330}.

\bibitem[{\citenamefont{Dvali et~al.}(2004{\natexlab{a}})\citenamefont{Dvali,
  Gruzinov, and Zaldarriaga}}]{Dvali:2003em}
\bibinfo{author}{\bibfnamefont{G.}~\bibnamefont{Dvali}},
  \bibinfo{author}{\bibfnamefont{A.}~\bibnamefont{Gruzinov}}, \bibnamefont{and}
  \bibinfo{author}{\bibfnamefont{M.}~\bibnamefont{Zaldarriaga}},
  \bibinfo{journal}{Phys. Rev.} \textbf{\bibinfo{volume}{D69}},
  \bibinfo{pages}{023505} (\bibinfo{year}{2004}{\natexlab{a}}),
  \eprint{astro-ph/0303591}.

\bibitem[{\citenamefont{Creminelli}(2003)}]{Creminelli:2003iq}
\bibinfo{author}{\bibfnamefont{P.}~\bibnamefont{Creminelli}},
  \bibinfo{journal}{JCAP} \textbf{\bibinfo{volume}{0310}}, \bibinfo{pages}{003}
  (\bibinfo{year}{2003}), \eprint{astro-ph/0306122}.

\bibitem[{\citenamefont{Alishahiha et~al.}(2004)\citenamefont{Alishahiha,
  Silverstein, and Tong}}]{Alishahiha:2004eh}
\bibinfo{author}{\bibfnamefont{M.}~\bibnamefont{Alishahiha}},
  \bibinfo{author}{\bibfnamefont{E.}~\bibnamefont{Silverstein}},
  \bibnamefont{and} \bibinfo{author}{\bibfnamefont{D.}~\bibnamefont{Tong}},
  \bibinfo{journal}{Phys. Rev.} \textbf{\bibinfo{volume}{D70}},
  \bibinfo{pages}{123505} (\bibinfo{year}{2004}), \eprint{hep-th/0404084}.

\bibitem[{\citenamefont{Gruzinov}(2005)}]{Gruzinov:2004jx}
\bibinfo{author}{\bibfnamefont{A.}~\bibnamefont{Gruzinov}},
  \bibinfo{journal}{Phys. Rev.} \textbf{\bibinfo{volume}{D71}},
  \bibinfo{pages}{027301} (\bibinfo{year}{2005}), \eprint{astro-ph/0406129}.

\bibitem[{\citenamefont{Enqvist
  et~al.}(2005{\natexlab{a}})\citenamefont{Enqvist, Jokinen, Mazumdar,
  Multamaki, and Vaihkonen}}]{Enqvist:2004ey}
\bibinfo{author}{\bibfnamefont{K.}~\bibnamefont{Enqvist}},
  \bibinfo{author}{\bibfnamefont{A.}~\bibnamefont{Jokinen}},
  \bibinfo{author}{\bibfnamefont{A.}~\bibnamefont{Mazumdar}},
  \bibinfo{author}{\bibfnamefont{T.}~\bibnamefont{Multamaki}},
  \bibnamefont{and}
  \bibinfo{author}{\bibfnamefont{A.}~\bibnamefont{Vaihkonen}},
  \bibinfo{journal}{Phys. Rev. Lett.} \textbf{\bibinfo{volume}{94}},
  \bibinfo{pages}{161301} (\bibinfo{year}{2005}{\natexlab{a}}),
  \eprint{astro-ph/0411394}.

\bibitem[{\citenamefont{Jokinen and Mazumdar}(2006)}]{Jokinen:2005by}
\bibinfo{author}{\bibfnamefont{A.}~\bibnamefont{Jokinen}} \bibnamefont{and}
  \bibinfo{author}{\bibfnamefont{A.}~\bibnamefont{Mazumdar}},
  \bibinfo{journal}{JCAP} \textbf{\bibinfo{volume}{0604}}, \bibinfo{pages}{003}
  (\bibinfo{year}{2006}), \eprint{astro-ph/0512368}.

\bibitem[{\citenamefont{Enqvist
  et~al.}(2005{\natexlab{b}})\citenamefont{Enqvist, Jokinen, Mazumdar,
  Multamaki, and Vaihkonen}}]{Enqvist:2005qu}
\bibinfo{author}{\bibfnamefont{K.}~\bibnamefont{Enqvist}},
  \bibinfo{author}{\bibfnamefont{A.}~\bibnamefont{Jokinen}},
  \bibinfo{author}{\bibfnamefont{A.}~\bibnamefont{Mazumdar}},
  \bibinfo{author}{\bibfnamefont{T.}~\bibnamefont{Multamaki}},
  \bibnamefont{and}
  \bibinfo{author}{\bibfnamefont{A.}~\bibnamefont{Vaihkonen}},
  \bibinfo{journal}{JCAP} \textbf{\bibinfo{volume}{0503}}, \bibinfo{pages}{010}
  (\bibinfo{year}{2005}{\natexlab{b}}), \eprint{hep-ph/0501076}.

\bibitem[{\citenamefont{Lyth}(2005)}]{Lyth:2005qk}
\bibinfo{author}{\bibfnamefont{D.~H.} \bibnamefont{Lyth}},
  \bibinfo{journal}{JCAP} \textbf{\bibinfo{volume}{0511}}, \bibinfo{pages}{006}
  (\bibinfo{year}{2005}), \eprint{astro-ph/0510443}.

\bibitem[{\citenamefont{Salem}(2005)}]{Salem:2005nd}
\bibinfo{author}{\bibfnamefont{M.~P.} \bibnamefont{Salem}},
  \bibinfo{journal}{Phys. Rev.} \textbf{\bibinfo{volume}{D72}},
  \bibinfo{pages}{123516} (\bibinfo{year}{2005}), \eprint{astro-ph/0511146}.

\bibitem[{\citenamefont{Seery and Lidsey}(2007)}]{Seery:2006js}
\bibinfo{author}{\bibfnamefont{D.}~\bibnamefont{Seery}} \bibnamefont{and}
  \bibinfo{author}{\bibfnamefont{J.~E.} \bibnamefont{Lidsey}},
  \bibinfo{journal}{JCAP} \textbf{\bibinfo{volume}{0701}}, \bibinfo{pages}{008}
  (\bibinfo{year}{2007}), \eprint{astro-ph/0611034}.

\bibitem[{\citenamefont{Seery and Lidsey}(2005{\natexlab{a}})}]{Seery:2005wm}
\bibinfo{author}{\bibfnamefont{D.}~\bibnamefont{Seery}} \bibnamefont{and}
  \bibinfo{author}{\bibfnamefont{J.~E.} \bibnamefont{Lidsey}},
  \bibinfo{journal}{JCAP} \textbf{\bibinfo{volume}{0506}}, \bibinfo{pages}{003}
  (\bibinfo{year}{2005}{\natexlab{a}}), \eprint{astro-ph/0503692}.

\bibitem[{\citenamefont{Seery and Lidsey}(2005{\natexlab{b}})}]{Seery:2005gb}
\bibinfo{author}{\bibfnamefont{D.}~\bibnamefont{Seery}} \bibnamefont{and}
  \bibinfo{author}{\bibfnamefont{J.~E.} \bibnamefont{Lidsey}},
  \bibinfo{journal}{JCAP} \textbf{\bibinfo{volume}{0509}}, \bibinfo{pages}{011}
  (\bibinfo{year}{2005}{\natexlab{b}}), \eprint{astro-ph/0506056}.

\bibitem[{\citenamefont{Sasaki et~al.}(2006)\citenamefont{Sasaki, Valiviita,
  and Wands}}]{Sasaki:2006kq}
\bibinfo{author}{\bibfnamefont{M.}~\bibnamefont{Sasaki}},
  \bibinfo{author}{\bibfnamefont{J.}~\bibnamefont{Valiviita}},
  \bibnamefont{and} \bibinfo{author}{\bibfnamefont{D.}~\bibnamefont{Wands}},
  \bibinfo{journal}{Phys. Rev.} \textbf{\bibinfo{volume}{D74}},
  \bibinfo{pages}{103003} (\bibinfo{year}{2006}), \eprint{astro-ph/0607627}.

\bibitem[{\citenamefont{Malik and Lyth}(2006)}]{Malik:2006pm}
\bibinfo{author}{\bibfnamefont{K.~A.} \bibnamefont{Malik}} \bibnamefont{and}
  \bibinfo{author}{\bibfnamefont{D.~H.} \bibnamefont{Lyth}},
  \bibinfo{journal}{JCAP} \textbf{\bibinfo{volume}{0609}}, \bibinfo{pages}{008}
  (\bibinfo{year}{2006}), \eprint{astro-ph/0604387}.

\bibitem[{\citenamefont{Barnaby and Cline}(2006)}]{Barnaby:2006cq}
\bibinfo{author}{\bibfnamefont{N.}~\bibnamefont{Barnaby}} \bibnamefont{and}
  \bibinfo{author}{\bibfnamefont{J.~M.} \bibnamefont{Cline}},
  \bibinfo{journal}{Phys. Rev.} \textbf{\bibinfo{volume}{D73}},
  \bibinfo{pages}{106012} (\bibinfo{year}{2006}), \eprint{astro-ph/0601481}.

\bibitem[{\citenamefont{Alabidi and Lyth}(2006)}]{Alabidi:2006wa}
\bibinfo{author}{\bibfnamefont{L.}~\bibnamefont{Alabidi}} \bibnamefont{and}
  \bibinfo{author}{\bibfnamefont{D.}~\bibnamefont{Lyth}},
  \bibinfo{journal}{JCAP} \textbf{\bibinfo{volume}{0608}}, \bibinfo{pages}{006}
  (\bibinfo{year}{2006}), \eprint{astro-ph/0604569}.

\bibitem[{\citenamefont{Chen et~al.}(2007{\natexlab{a}})\citenamefont{Chen,
  Huang, Kachru, and Shiu}}]{Chen:2006nt}
\bibinfo{author}{\bibfnamefont{X.}~\bibnamefont{Chen}},
  \bibinfo{author}{\bibfnamefont{M.-x.} \bibnamefont{Huang}},
  \bibinfo{author}{\bibfnamefont{S.}~\bibnamefont{Kachru}}, \bibnamefont{and}
  \bibinfo{author}{\bibfnamefont{G.}~\bibnamefont{Shiu}},
  \bibinfo{journal}{JCAP} \textbf{\bibinfo{volume}{0701}}, \bibinfo{pages}{002}
  (\bibinfo{year}{2007}{\natexlab{a}}), \eprint{hep-th/0605045}.

\bibitem[{\citenamefont{Chen et~al.}(2006)\citenamefont{Chen, Huang, and
  Shiu}}]{Huang:2006eh}
\bibinfo{author}{\bibfnamefont{X.}~\bibnamefont{Chen}},
  \bibinfo{author}{\bibfnamefont{M.-x.} \bibnamefont{Huang}}, \bibnamefont{and}
  \bibinfo{author}{\bibfnamefont{G.}~\bibnamefont{Shiu}},
  \bibinfo{journal}{Phys. Rev.} \textbf{\bibinfo{volume}{D74}},
  \bibinfo{pages}{121301} (\bibinfo{year}{2006}), \eprint{hep-th/0610235}.

\bibitem[{\citenamefont{Chen et~al.}(2007{\natexlab{b}})\citenamefont{Chen,
  Easther, and Lim}}]{Chen:2006xjb}
\bibinfo{author}{\bibfnamefont{X.}~\bibnamefont{Chen}},
  \bibinfo{author}{\bibfnamefont{R.}~\bibnamefont{Easther}}, \bibnamefont{and}
  \bibinfo{author}{\bibfnamefont{E.~A.} \bibnamefont{Lim}},
  \bibinfo{journal}{JCAP} \textbf{\bibinfo{volume}{0706}}, \bibinfo{pages}{023}
  (\bibinfo{year}{2007}{\natexlab{b}}), \eprint{astro-ph/0611645}.

\bibitem[{\citenamefont{Alabidi}(2006)}]{Alabidi:2006hg}
\bibinfo{author}{\bibfnamefont{L.}~\bibnamefont{Alabidi}},
  \bibinfo{journal}{JCAP} \textbf{\bibinfo{volume}{0610}}, \bibinfo{pages}{015}
  (\bibinfo{year}{2006}), \eprint{astro-ph/0604611}.

\bibitem[{\citenamefont{Byrnes et~al.}(2006)\citenamefont{Byrnes, Sasaki, and
  Wands}}]{Byrnes:2006vq}
\bibinfo{author}{\bibfnamefont{C.~T.} \bibnamefont{Byrnes}},
  \bibinfo{author}{\bibfnamefont{M.}~\bibnamefont{Sasaki}}, \bibnamefont{and}
  \bibinfo{author}{\bibfnamefont{D.}~\bibnamefont{Wands}},
  \bibinfo{journal}{Phys. Rev.} \textbf{\bibinfo{volume}{D74}},
  \bibinfo{pages}{123519} (\bibinfo{year}{2006}), \eprint{astro-ph/0611075}.

\bibitem[{\citenamefont{Arroja and Koyama}(2008)}]{Arroja:2008ga}
\bibinfo{author}{\bibfnamefont{F.}~\bibnamefont{Arroja}} \bibnamefont{and}
  \bibinfo{author}{\bibfnamefont{K.}~\bibnamefont{Koyama}},
  \bibinfo{journal}{Phys. Rev.} \textbf{\bibinfo{volume}{D77}},
  \bibinfo{pages}{083517} (\bibinfo{year}{2008}), \eprint{0802.1167}.

\bibitem[{\citenamefont{Arroja et~al.}(2008)\citenamefont{Arroja, Mizuno, and
  Koyama}}]{Arroja:2008yy}
\bibinfo{author}{\bibfnamefont{F.}~\bibnamefont{Arroja}},
  \bibinfo{author}{\bibfnamefont{S.}~\bibnamefont{Mizuno}}, \bibnamefont{and}
  \bibinfo{author}{\bibfnamefont{K.}~\bibnamefont{Koyama}},
  \bibinfo{journal}{JCAP} \textbf{\bibinfo{volume}{0808}}, \bibinfo{pages}{015}
  (\bibinfo{year}{2008}), \eprint{0806.0619}.

\bibitem[{\citenamefont{Langlois
  et~al.}(2008{\natexlab{a}})\citenamefont{Langlois, Renaux-Petel, Steer, and
  Tanaka}}]{Langlois:2008wt}
\bibinfo{author}{\bibfnamefont{D.}~\bibnamefont{Langlois}},
  \bibinfo{author}{\bibfnamefont{S.}~\bibnamefont{Renaux-Petel}},
  \bibinfo{author}{\bibfnamefont{D.~A.} \bibnamefont{Steer}}, \bibnamefont{and}
  \bibinfo{author}{\bibfnamefont{T.}~\bibnamefont{Tanaka}},
  \bibinfo{journal}{Phys. Rev. Lett.} \textbf{\bibinfo{volume}{101}},
  \bibinfo{pages}{061301} (\bibinfo{year}{2008}{\natexlab{a}}),
  \eprint{0804.3139}.

\bibitem[{\citenamefont{Langlois
  et~al.}(2008{\natexlab{b}})\citenamefont{Langlois, Renaux-Petel, Steer, and
  Tanaka}}]{Langlois:2008qf}
\bibinfo{author}{\bibfnamefont{D.}~\bibnamefont{Langlois}},
  \bibinfo{author}{\bibfnamefont{S.}~\bibnamefont{Renaux-Petel}},
  \bibinfo{author}{\bibfnamefont{D.~A.} \bibnamefont{Steer}}, \bibnamefont{and}
  \bibinfo{author}{\bibfnamefont{T.}~\bibnamefont{Tanaka}},
  \bibinfo{journal}{Phys. Rev.} \textbf{\bibinfo{volume}{D78}},
  \bibinfo{pages}{063523} (\bibinfo{year}{2008}{\natexlab{b}}),
  \eprint{0806.0336}.

\bibitem[{\citenamefont{Sasaki}(2008)}]{Sasaki:2008uc}
\bibinfo{author}{\bibfnamefont{M.}~\bibnamefont{Sasaki}},
  \bibinfo{journal}{Prog. Theor. Phys.} \textbf{\bibinfo{volume}{120}},
  \bibinfo{pages}{159} (\bibinfo{year}{2008}), \eprint{0805.0974}.

\bibitem[{\citenamefont{Byrnes et~al.}(2008)\citenamefont{Byrnes, Choi, and
  Hall}}]{Byrnes:2008wi}
\bibinfo{author}{\bibfnamefont{C.~T.} \bibnamefont{Byrnes}},
  \bibinfo{author}{\bibfnamefont{K.-Y.} \bibnamefont{Choi}}, \bibnamefont{and}
  \bibinfo{author}{\bibfnamefont{L.~M.~H.} \bibnamefont{Hall}},
  \bibinfo{journal}{JCAP} \textbf{\bibinfo{volume}{0810}}, \bibinfo{pages}{008}
  (\bibinfo{year}{2008}), \eprint{0807.1101}.

\bibitem[{\citenamefont{Byrnes et~al.}(2009)\citenamefont{Byrnes, Choi, and
  Hall}}]{Byrnes:2008zy}
\bibinfo{author}{\bibfnamefont{C.~T.} \bibnamefont{Byrnes}},
  \bibinfo{author}{\bibfnamefont{K.-Y.} \bibnamefont{Choi}}, \bibnamefont{and}
  \bibinfo{author}{\bibfnamefont{L.~M.~H.} \bibnamefont{Hall}},
  \bibinfo{journal}{JCAP} \textbf{\bibinfo{volume}{0902}}, \bibinfo{pages}{017}
  (\bibinfo{year}{2009}), \eprint{0812.0807}.

\bibitem[{\citenamefont{Yokoyama et~al.}(2007)\citenamefont{Yokoyama, Suyama,
  and Tanaka}}]{Yokoyama:2007uu}
\bibinfo{author}{\bibfnamefont{S.}~\bibnamefont{Yokoyama}},
  \bibinfo{author}{\bibfnamefont{T.}~\bibnamefont{Suyama}}, \bibnamefont{and}
  \bibinfo{author}{\bibfnamefont{T.}~\bibnamefont{Tanaka}},
  \bibinfo{journal}{JCAP} \textbf{\bibinfo{volume}{0707}}, \bibinfo{pages}{013}
  (\bibinfo{year}{2007}), \eprint{0705.3178}.

\bibitem[{\citenamefont{Yokoyama et~al.}(2008)\citenamefont{Yokoyama, Suyama,
  and Tanaka}}]{Yokoyama:2007dw}
\bibinfo{author}{\bibfnamefont{S.}~\bibnamefont{Yokoyama}},
  \bibinfo{author}{\bibfnamefont{T.}~\bibnamefont{Suyama}}, \bibnamefont{and}
  \bibinfo{author}{\bibfnamefont{T.}~\bibnamefont{Tanaka}},
  \bibinfo{journal}{Phys. Rev.} \textbf{\bibinfo{volume}{D77}},
  \bibinfo{pages}{083511} (\bibinfo{year}{2008}), \eprint{0711.2920}.

\bibitem[{\citenamefont{Dutta et~al.}(2008)\citenamefont{Dutta, Leblond, and
  Kumar}}]{Dutta:2008if}
\bibinfo{author}{\bibfnamefont{B.}~\bibnamefont{Dutta}},
  \bibinfo{author}{\bibfnamefont{L.}~\bibnamefont{Leblond}}, \bibnamefont{and}
  \bibinfo{author}{\bibfnamefont{J.}~\bibnamefont{Kumar}},
  \bibinfo{journal}{Phys. Rev.} \textbf{\bibinfo{volume}{D78}},
  \bibinfo{pages}{083522} (\bibinfo{year}{2008}), \eprint{0805.1229}.

\bibitem[{\citenamefont{Naruko and Sasaki}(2009)}]{Naruko:2008sq}
\bibinfo{author}{\bibfnamefont{A.}~\bibnamefont{Naruko}} \bibnamefont{and}
  \bibinfo{author}{\bibfnamefont{M.}~\bibnamefont{Sasaki}},
  \bibinfo{journal}{Prog. Theor. Phys.} \textbf{\bibinfo{volume}{121}},
  \bibinfo{pages}{193} (\bibinfo{year}{2009}), \eprint{0807.0180}.

\bibitem[{\citenamefont{Suyama and Yamaguchi}(2008)}]{Suyama:2007bg}
\bibinfo{author}{\bibfnamefont{T.}~\bibnamefont{Suyama}} \bibnamefont{and}
  \bibinfo{author}{\bibfnamefont{M.}~\bibnamefont{Yamaguchi}},
  \bibinfo{journal}{Phys. Rev.} \textbf{\bibinfo{volume}{D77}},
  \bibinfo{pages}{023505} (\bibinfo{year}{2008}), \eprint{0709.2545}.

\bibitem[{\citenamefont{Suyama and Takahashi}(2008)}]{Suyama:2008nt}
\bibinfo{author}{\bibfnamefont{T.}~\bibnamefont{Suyama}} \bibnamefont{and}
  \bibinfo{author}{\bibfnamefont{F.}~\bibnamefont{Takahashi}},
  \bibinfo{journal}{JCAP} \textbf{\bibinfo{volume}{0809}}, \bibinfo{pages}{007}
  (\bibinfo{year}{2008}), \eprint{0804.0425}.

\bibitem[{\citenamefont{Gao}(2008)}]{Gao:2008dt}
\bibinfo{author}{\bibfnamefont{X.}~\bibnamefont{Gao}}, \bibinfo{journal}{JCAP}
  \textbf{\bibinfo{volume}{0806}}, \bibinfo{pages}{029} (\bibinfo{year}{2008}),
  \eprint{0804.1055}.

\bibitem[{\citenamefont{Cogollo et~al.}(2008)\citenamefont{Cogollo, Rodriguez,
  and Valenzuela-Toledo}}]{Cogollo:2008bi}
\bibinfo{author}{\bibfnamefont{H.~R.~S.} \bibnamefont{Cogollo}},
  \bibinfo{author}{\bibfnamefont{Y.}~\bibnamefont{Rodriguez}},
  \bibnamefont{and} \bibinfo{author}{\bibfnamefont{C.~A.}
  \bibnamefont{Valenzuela-Toledo}}, \bibinfo{journal}{JCAP}
  \textbf{\bibinfo{volume}{0808}}, \bibinfo{pages}{029} (\bibinfo{year}{2008}),
  \eprint{0806.1546}.

\bibitem[{\citenamefont{Rodriguez and
  Valenzuela-Toledo}(2008)}]{Rodriguez:2008hy}
\bibinfo{author}{\bibfnamefont{Y.}~\bibnamefont{Rodriguez}} \bibnamefont{and}
  \bibinfo{author}{\bibfnamefont{C.~A.} \bibnamefont{Valenzuela-Toledo}}
  (\bibinfo{year}{2008}), \eprint{0811.4092}.

\bibitem[{\citenamefont{Ichikawa et~al.}(2008)\citenamefont{Ichikawa, Suyama,
  Takahashi, and Yamaguchi}}]{Ichikawa:2008iq}
\bibinfo{author}{\bibfnamefont{K.}~\bibnamefont{Ichikawa}},
  \bibinfo{author}{\bibfnamefont{T.}~\bibnamefont{Suyama}},
  \bibinfo{author}{\bibfnamefont{T.}~\bibnamefont{Takahashi}},
  \bibnamefont{and}
  \bibinfo{author}{\bibfnamefont{M.}~\bibnamefont{Yamaguchi}},
  \bibinfo{journal}{Phys. Rev.} \textbf{\bibinfo{volume}{D78}},
  \bibinfo{pages}{023513} (\bibinfo{year}{2008}), \eprint{0802.4138}.

\bibitem[{\citenamefont{Byrnes}(2009)}]{Byrnes:2008zz}
\bibinfo{author}{\bibfnamefont{C.~T.} \bibnamefont{Byrnes}},
  \bibinfo{journal}{JCAP} \textbf{\bibinfo{volume}{0901}}, \bibinfo{pages}{011}
  (\bibinfo{year}{2009}), \eprint{0810.3913}.

\bibitem[{\citenamefont{Li et~al.}(2009)\citenamefont{Li, Cai, and
  Piao}}]{Li:2008fma}
\bibinfo{author}{\bibfnamefont{S.}~\bibnamefont{Li}},
  \bibinfo{author}{\bibfnamefont{Y.-F.} \bibnamefont{Cai}}, \bibnamefont{and}
  \bibinfo{author}{\bibfnamefont{Y.-S.} \bibnamefont{Piao}},
  \bibinfo{journal}{Phys. Lett.} \textbf{\bibinfo{volume}{B671}},
  \bibinfo{pages}{423} (\bibinfo{year}{2009}), \eprint{0806.2363}.

\bibitem[{\citenamefont{Langlois
  et~al.}(2008{\natexlab{c}})\citenamefont{Langlois, Vernizzi, and
  Wands}}]{Langlois:2008vk}
\bibinfo{author}{\bibfnamefont{D.}~\bibnamefont{Langlois}},
  \bibinfo{author}{\bibfnamefont{F.}~\bibnamefont{Vernizzi}}, \bibnamefont{and}
  \bibinfo{author}{\bibfnamefont{D.}~\bibnamefont{Wands}},
  \bibinfo{journal}{JCAP} \textbf{\bibinfo{volume}{0812}}, \bibinfo{pages}{004}
  (\bibinfo{year}{2008}{\natexlab{c}}), \eprint{0809.4646}.

\bibitem[{\citenamefont{Hikage et~al.}(2008)\citenamefont{Hikage, Koyama,
  Matsubara, Takahashi, and Yamaguchi}}]{Hikage:2008sk}
\bibinfo{author}{\bibfnamefont{C.}~\bibnamefont{Hikage}},
  \bibinfo{author}{\bibfnamefont{K.}~\bibnamefont{Koyama}},
  \bibinfo{author}{\bibfnamefont{T.}~\bibnamefont{Matsubara}},
  \bibinfo{author}{\bibfnamefont{T.}~\bibnamefont{Takahashi}},
  \bibnamefont{and} \bibinfo{author}{\bibfnamefont{M.}~\bibnamefont{Yamaguchi}}
  (\bibinfo{year}{2008}), \eprint{0812.3500}.

\bibitem[{\citenamefont{Kawasaki et~al.}(2008)\citenamefont{Kawasaki, Nakayama,
  Sekiguchi, Suyama, and Takahashi}}]{Kawasaki:2008sn}
\bibinfo{author}{\bibfnamefont{M.}~\bibnamefont{Kawasaki}},
  \bibinfo{author}{\bibfnamefont{K.}~\bibnamefont{Nakayama}},
  \bibinfo{author}{\bibfnamefont{T.}~\bibnamefont{Sekiguchi}},
  \bibinfo{author}{\bibfnamefont{T.}~\bibnamefont{Suyama}}, \bibnamefont{and}
  \bibinfo{author}{\bibfnamefont{F.}~\bibnamefont{Takahashi}},
  \bibinfo{journal}{JCAP} \textbf{\bibinfo{volume}{0811}}, \bibinfo{pages}{019}
  (\bibinfo{year}{2008}), \eprint{0808.0009}.

\bibitem[{\citenamefont{Creminelli and Senatore}(2007)}]{Creminelli:2007aq}
\bibinfo{author}{\bibfnamefont{P.}~\bibnamefont{Creminelli}} \bibnamefont{and}
  \bibinfo{author}{\bibfnamefont{L.}~\bibnamefont{Senatore}},
  \bibinfo{journal}{JCAP} \textbf{\bibinfo{volume}{0711}}, \bibinfo{pages}{010}
  (\bibinfo{year}{2007}), \eprint{hep-th/0702165}.

\bibitem[{\citenamefont{Koyama et~al.}(2007)\citenamefont{Koyama, Mizuno,
  Vernizzi, and Wands}}]{Koyama:2007if}
\bibinfo{author}{\bibfnamefont{K.}~\bibnamefont{Koyama}},
  \bibinfo{author}{\bibfnamefont{S.}~\bibnamefont{Mizuno}},
  \bibinfo{author}{\bibfnamefont{F.}~\bibnamefont{Vernizzi}}, \bibnamefont{and}
  \bibinfo{author}{\bibfnamefont{D.}~\bibnamefont{Wands}},
  \bibinfo{journal}{JCAP} \textbf{\bibinfo{volume}{0711}}, \bibinfo{pages}{024}
  (\bibinfo{year}{2007}), \eprint{0708.4321}.

\bibitem[{\citenamefont{Buchbinder et~al.}(2008)\citenamefont{Buchbinder,
  Khoury, and Ovrut}}]{Buchbinder:2007at}
\bibinfo{author}{\bibfnamefont{E.~I.} \bibnamefont{Buchbinder}},
  \bibinfo{author}{\bibfnamefont{J.}~\bibnamefont{Khoury}}, \bibnamefont{and}
  \bibinfo{author}{\bibfnamefont{B.~A.} \bibnamefont{Ovrut}},
  \bibinfo{journal}{Phys. Rev. Lett.} \textbf{\bibinfo{volume}{100}},
  \bibinfo{pages}{171302} (\bibinfo{year}{2008}), \eprint{0710.5172}.

\bibitem[{\citenamefont{Lehners and
  Steinhardt}(2008{\natexlab{a}})}]{Lehners:2007wc}
\bibinfo{author}{\bibfnamefont{J.-L.} \bibnamefont{Lehners}} \bibnamefont{and}
  \bibinfo{author}{\bibfnamefont{P.~J.} \bibnamefont{Steinhardt}},
  \bibinfo{journal}{Phys. Rev.} \textbf{\bibinfo{volume}{D77}},
  \bibinfo{pages}{063533} (\bibinfo{year}{2008}{\natexlab{a}}),
  \eprint{0712.3779}.

\bibitem[{\citenamefont{Lehners and
  Steinhardt}(2008{\natexlab{b}})}]{Lehners:2008my}
\bibinfo{author}{\bibfnamefont{J.-L.} \bibnamefont{Lehners}} \bibnamefont{and}
  \bibinfo{author}{\bibfnamefont{P.~J.} \bibnamefont{Steinhardt}},
  \bibinfo{journal}{Phys. Rev.} \textbf{\bibinfo{volume}{D78}},
  \bibinfo{pages}{023506} (\bibinfo{year}{2008}{\natexlab{b}}),
  \eprint{0804.1293}.

\bibitem[{\citenamefont{Misra and Shukla}(2009)}]{Misra:2008tx}
\bibinfo{author}{\bibfnamefont{A.}~\bibnamefont{Misra}} \bibnamefont{and}
  \bibinfo{author}{\bibfnamefont{P.}~\bibnamefont{Shukla}},
  \bibinfo{journal}{Nucl. Phys.} \textbf{\bibinfo{volume}{B810}},
  \bibinfo{pages}{174} (\bibinfo{year}{2009}), \eprint{0807.0996}.

\bibitem[{\citenamefont{Huang}(2009)}]{Huang:2009vk}
\bibinfo{author}{\bibfnamefont{Q.-G.} \bibnamefont{Huang}},
  \bibinfo{journal}{JCAP} \textbf{\bibinfo{volume}{0906}}, \bibinfo{pages}{035}
  (\bibinfo{year}{2009}), \eprint{0904.2649}.

\bibitem[{\citenamefont{Khoury and Piazza}(2009)}]{Khoury:2008wj}
\bibinfo{author}{\bibfnamefont{J.}~\bibnamefont{Khoury}} \bibnamefont{and}
  \bibinfo{author}{\bibfnamefont{F.}~\bibnamefont{Piazza}},
  \bibinfo{journal}{JCAP} \textbf{\bibinfo{volume}{0907}}, \bibinfo{pages}{026}
  (\bibinfo{year}{2009}), \eprint{0811.3633}.

\bibitem[{\citenamefont{Verde et~al.}(2000)\citenamefont{Verde, Wang, Heavens,
  and Kamionkowski}}]{Verde:1999ij}
\bibinfo{author}{\bibfnamefont{L.}~\bibnamefont{Verde}},
  \bibinfo{author}{\bibfnamefont{L.-M.} \bibnamefont{Wang}},
  \bibinfo{author}{\bibfnamefont{A.}~\bibnamefont{Heavens}}, \bibnamefont{and}
  \bibinfo{author}{\bibfnamefont{M.}~\bibnamefont{Kamionkowski}},
  \bibinfo{journal}{Mon. Not. Roy. Astron. Soc.}
  \textbf{\bibinfo{volume}{313}}, \bibinfo{pages}{L141} (\bibinfo{year}{2000}),
  \eprint{astro-ph/9906301}.

\bibitem[{\citenamefont{Wang and Kamionkowski}(2000)}]{Wang:1999vf}
\bibinfo{author}{\bibfnamefont{L.-M.} \bibnamefont{Wang}} \bibnamefont{and}
  \bibinfo{author}{\bibfnamefont{M.}~\bibnamefont{Kamionkowski}},
  \bibinfo{journal}{Phys. Rev.} \textbf{\bibinfo{volume}{D61}},
  \bibinfo{pages}{063504} (\bibinfo{year}{2000}), \eprint{astro-ph/9907431}.

\bibitem[{\citenamefont{Komatsu and Spergel}(2001)}]{Komatsu:2001rj}
\bibinfo{author}{\bibfnamefont{E.}~\bibnamefont{Komatsu}} \bibnamefont{and}
  \bibinfo{author}{\bibfnamefont{D.~N.} \bibnamefont{Spergel}},
  \bibinfo{journal}{Phys. Rev.} \textbf{\bibinfo{volume}{D63}},
  \bibinfo{pages}{063002} (\bibinfo{year}{2001}), \eprint{astro-ph/0005036}.

\bibitem[{\citenamefont{Starobinsky}(1985)}]{Starobinsky:1986fxa}
\bibinfo{author}{\bibfnamefont{A.~A.} \bibnamefont{Starobinsky}},
  \bibinfo{journal}{JETP Lett.} \textbf{\bibinfo{volume}{42}},
  \bibinfo{pages}{152} (\bibinfo{year}{1985}).

\bibitem[{\citenamefont{Sasaki and Stewart}(1996)}]{Sasaki:1995aw}
\bibinfo{author}{\bibfnamefont{M.}~\bibnamefont{Sasaki}} \bibnamefont{and}
  \bibinfo{author}{\bibfnamefont{E.~D.} \bibnamefont{Stewart}},
  \bibinfo{journal}{Prog. Theor. Phys.} \textbf{\bibinfo{volume}{95}},
  \bibinfo{pages}{71} (\bibinfo{year}{1996}), \eprint{astro-ph/9507001}.

\bibitem[{\citenamefont{Lyth et~al.}(2005)\citenamefont{Lyth, Malik, and
  Sasaki}}]{Lyth:2004gb}
\bibinfo{author}{\bibfnamefont{D.~H.} \bibnamefont{Lyth}},
  \bibinfo{author}{\bibfnamefont{K.~A.} \bibnamefont{Malik}}, \bibnamefont{and}
  \bibinfo{author}{\bibfnamefont{M.}~\bibnamefont{Sasaki}},
  \bibinfo{journal}{JCAP} \textbf{\bibinfo{volume}{0505}}, \bibinfo{pages}{004}
  (\bibinfo{year}{2005}), \eprint{astro-ph/0411220}.

\bibitem[{\citenamefont{Lyth and Rodriguez}(2005)}]{Lyth:2005fi}
\bibinfo{author}{\bibfnamefont{D.~H.} \bibnamefont{Lyth}} \bibnamefont{and}
  \bibinfo{author}{\bibfnamefont{Y.}~\bibnamefont{Rodriguez}},
  \bibinfo{journal}{Phys. Rev. Lett.} \textbf{\bibinfo{volume}{95}},
  \bibinfo{pages}{121302} (\bibinfo{year}{2005}), \eprint{astro-ph/0504045}.

\bibitem[{\citenamefont{Tanaka et~al.}(2010)\citenamefont{Tanaka, Suyama, and
  Yokoyama}}]{Tanaka:2010km}
\bibinfo{author}{\bibfnamefont{T.}~\bibnamefont{Tanaka}},
  \bibinfo{author}{\bibfnamefont{T.}~\bibnamefont{Suyama}}, \bibnamefont{and}
  \bibinfo{author}{\bibfnamefont{S.}~\bibnamefont{Yokoyama}}
  (\bibinfo{year}{2010}), \eprint{1003.5057}.

\bibitem[{\citenamefont{Salopek and Bond}(1990)}]{Salopek:1990jq}
\bibinfo{author}{\bibfnamefont{D.~S.} \bibnamefont{Salopek}} \bibnamefont{and}
  \bibinfo{author}{\bibfnamefont{J.~R.} \bibnamefont{Bond}},
  \bibinfo{journal}{Phys. Rev.} \textbf{\bibinfo{volume}{D42}},
  \bibinfo{pages}{3936} (\bibinfo{year}{1990}).

\bibitem[{\citenamefont{Sasaki and Tanaka}(1998)}]{Sasaki:1998ug}
\bibinfo{author}{\bibfnamefont{M.}~\bibnamefont{Sasaki}} \bibnamefont{and}
  \bibinfo{author}{\bibfnamefont{T.}~\bibnamefont{Tanaka}},
  \bibinfo{journal}{Prog. Theor. Phys.} \textbf{\bibinfo{volume}{99}},
  \bibinfo{pages}{763} (\bibinfo{year}{1998}), \eprint{gr-qc/9801017}.

\bibitem[{\citenamefont{Wands et~al.}(2000)\citenamefont{Wands, Malik, Lyth,
  and Liddle}}]{Wands:2000dp}
\bibinfo{author}{\bibfnamefont{D.}~\bibnamefont{Wands}},
  \bibinfo{author}{\bibfnamefont{K.~A.} \bibnamefont{Malik}},
  \bibinfo{author}{\bibfnamefont{D.~H.} \bibnamefont{Lyth}}, \bibnamefont{and}
  \bibinfo{author}{\bibfnamefont{A.~R.} \bibnamefont{Liddle}},
  \bibinfo{journal}{Phys. Rev.} \textbf{\bibinfo{volume}{D62}},
  \bibinfo{pages}{043527} (\bibinfo{year}{2000}), \eprint{astro-ph/0003278}.

\bibitem[{\citenamefont{Byrnes et~al.}(2007)\citenamefont{Byrnes, Koyama,
  Sasaki, and Wands}}]{Byrnes:2007tm}
\bibinfo{author}{\bibfnamefont{C.~T.} \bibnamefont{Byrnes}},
  \bibinfo{author}{\bibfnamefont{K.}~\bibnamefont{Koyama}},
  \bibinfo{author}{\bibfnamefont{M.}~\bibnamefont{Sasaki}}, \bibnamefont{and}
  \bibinfo{author}{\bibfnamefont{D.}~\bibnamefont{Wands}},
  \bibinfo{journal}{JCAP} \textbf{\bibinfo{volume}{0711}}, \bibinfo{pages}{027}
  (\bibinfo{year}{2007}), \eprint{0705.4096}.

\bibitem[{\citenamefont{Weinberg}(2005)}]{Weinberg:2005vy}
\bibinfo{author}{\bibfnamefont{S.}~\bibnamefont{Weinberg}},
  \bibinfo{journal}{Phys. Rev.} \textbf{\bibinfo{volume}{D72}},
  \bibinfo{pages}{043514} (\bibinfo{year}{2005}), \eprint{hep-th/0506236}.

\bibitem[{\citenamefont{Peskin and Schroeder}()}]{Peskin:1995ev}
\bibinfo{author}{\bibfnamefont{M.~E.} \bibnamefont{Peskin}} \bibnamefont{and}
  \bibinfo{author}{\bibfnamefont{D.~V.} \bibnamefont{Schroeder}} (????),
  \bibinfo{note}{reading, USA: Addison-Wesley (1995) 842 p}.

\bibitem[{\citenamefont{Garriga and Mukhanov}(1999)}]{Garriga:1999vw}
\bibinfo{author}{\bibfnamefont{J.}~\bibnamefont{Garriga}} \bibnamefont{and}
  \bibinfo{author}{\bibfnamefont{V.~F.} \bibnamefont{Mukhanov}},
  \bibinfo{journal}{Phys. Lett.} \textbf{\bibinfo{volume}{B458}},
  \bibinfo{pages}{219} (\bibinfo{year}{1999}), \eprint{hep-th/9904176}.

\bibitem[{\citenamefont{Christopherson and
  Malik}(2008)}]{Christopherson:2008ry}
\bibinfo{author}{\bibfnamefont{A.~J.} \bibnamefont{Christopherson}}
  \bibnamefont{and} \bibinfo{author}{\bibfnamefont{K.~A.} \bibnamefont{Malik}}
  (\bibinfo{year}{2008}), \eprint{0809.3518}.

\bibitem[{\citenamefont{Arroja and Sasaki}(2010)}]{Arroja:2010wy}
\bibinfo{author}{\bibfnamefont{F.}~\bibnamefont{Arroja}} \bibnamefont{and}
  \bibinfo{author}{\bibfnamefont{M.}~\bibnamefont{Sasaki}}
  (\bibinfo{year}{2010}), \eprint{1002.1376}.

\bibitem[{\citenamefont{Arnowitt et~al.}(1960)\citenamefont{Arnowitt, Deser,
  and Misner}}]{Arnowitt:1960es}
\bibinfo{author}{\bibfnamefont{R.}~\bibnamefont{Arnowitt}},
  \bibinfo{author}{\bibfnamefont{S.}~\bibnamefont{Deser}}, \bibnamefont{and}
  \bibinfo{author}{\bibfnamefont{C.~W.} \bibnamefont{Misner}},
  \bibinfo{journal}{Phys. Rev.} \textbf{\bibinfo{volume}{117}},
  \bibinfo{pages}{1595} (\bibinfo{year}{1960}).

\bibitem[{\citenamefont{Seery et~al.}(2007)\citenamefont{Seery, Lidsey, and
  Sloth}}]{Seery:2006vu}
\bibinfo{author}{\bibfnamefont{D.}~\bibnamefont{Seery}},
  \bibinfo{author}{\bibfnamefont{J.~E.} \bibnamefont{Lidsey}},
  \bibnamefont{and} \bibinfo{author}{\bibfnamefont{M.~S.} \bibnamefont{Sloth}},
  \bibinfo{journal}{JCAP} \textbf{\bibinfo{volume}{0701}}, \bibinfo{pages}{027}
  (\bibinfo{year}{2007}), \eprint{astro-ph/0610210}.

\bibitem[{\citenamefont{Babich et~al.}(2004)\citenamefont{Babich, Creminelli,
  and Zaldarriaga}}]{Babich:2004gb}
\bibinfo{author}{\bibfnamefont{D.}~\bibnamefont{Babich}},
  \bibinfo{author}{\bibfnamefont{P.}~\bibnamefont{Creminelli}},
  \bibnamefont{and}
  \bibinfo{author}{\bibfnamefont{M.}~\bibnamefont{Zaldarriaga}},
  \bibinfo{journal}{JCAP} \textbf{\bibinfo{volume}{0408}}, \bibinfo{pages}{009}
  (\bibinfo{year}{2004}), \eprint{astro-ph/0405356}.

\bibitem[{\citenamefont{Creminelli et~al.}(2006)\citenamefont{Creminelli,
  Nicolis, Senatore, Tegmark, and Zaldarriaga}}]{Creminelli:2005hu}
\bibinfo{author}{\bibfnamefont{P.}~\bibnamefont{Creminelli}},
  \bibinfo{author}{\bibfnamefont{A.}~\bibnamefont{Nicolis}},
  \bibinfo{author}{\bibfnamefont{L.}~\bibnamefont{Senatore}},
  \bibinfo{author}{\bibfnamefont{M.}~\bibnamefont{Tegmark}}, \bibnamefont{and}
  \bibinfo{author}{\bibfnamefont{M.}~\bibnamefont{Zaldarriaga}},
  \bibinfo{journal}{JCAP} \textbf{\bibinfo{volume}{0605}}, \bibinfo{pages}{004}
  (\bibinfo{year}{2006}), \eprint{astro-ph/0509029}.

\bibitem[{\citenamefont{Komatsu et~al.}(2010)}]{Komatsu:2010fb}
\bibinfo{author}{\bibfnamefont{E.}~\bibnamefont{Komatsu}} \bibnamefont{et~al.}
  (\bibinfo{year}{2010}), \eprint{1001.4538}.

\bibitem[{\citenamefont{Lyth et~al.}(2003)\citenamefont{Lyth, Ungarelli, and
  Wands}}]{Lyth:2002my}
\bibinfo{author}{\bibfnamefont{D.~H.} \bibnamefont{Lyth}},
  \bibinfo{author}{\bibfnamefont{C.}~\bibnamefont{Ungarelli}},
  \bibnamefont{and} \bibinfo{author}{\bibfnamefont{D.}~\bibnamefont{Wands}},
  \bibinfo{journal}{Phys. Rev.} \textbf{\bibinfo{volume}{D67}},
  \bibinfo{pages}{023503} (\bibinfo{year}{2003}), \eprint{astro-ph/0208055}.

\bibitem[{\citenamefont{Dvali et~al.}(2004{\natexlab{b}})\citenamefont{Dvali,
  Gruzinov, and Zaldarriaga}}]{Dvali:2003ar}
\bibinfo{author}{\bibfnamefont{G.}~\bibnamefont{Dvali}},
  \bibinfo{author}{\bibfnamefont{A.}~\bibnamefont{Gruzinov}}, \bibnamefont{and}
  \bibinfo{author}{\bibfnamefont{M.}~\bibnamefont{Zaldarriaga}},
  \bibinfo{journal}{Phys. Rev.} \textbf{\bibinfo{volume}{D69}},
  \bibinfo{pages}{083505} (\bibinfo{year}{2004}{\natexlab{b}}),
  \eprint{astro-ph/0305548}.

\bibitem[{\citenamefont{Dalal et~al.}(2008)\citenamefont{Dalal, Dore, Huterer,
  and Shirokov}}]{Dalal:2007cu}
\bibinfo{author}{\bibfnamefont{N.}~\bibnamefont{Dalal}},
  \bibinfo{author}{\bibfnamefont{O.}~\bibnamefont{Dore}},
  \bibinfo{author}{\bibfnamefont{D.}~\bibnamefont{Huterer}}, \bibnamefont{and}
  \bibinfo{author}{\bibfnamefont{A.}~\bibnamefont{Shirokov}},
  \bibinfo{journal}{Phys. Rev.} \textbf{\bibinfo{volume}{D77}},
  \bibinfo{pages}{123514} (\bibinfo{year}{2008}), \eprint{0710.4560}.

\bibitem[{\citenamefont{Taruya et~al.}(2008)\citenamefont{Taruya, Koyama, and
  Matsubara}}]{Taruya:2008pg}
\bibinfo{author}{\bibfnamefont{A.}~\bibnamefont{Taruya}},
  \bibinfo{author}{\bibfnamefont{K.}~\bibnamefont{Koyama}}, \bibnamefont{and}
  \bibinfo{author}{\bibfnamefont{T.}~\bibnamefont{Matsubara}},
  \bibinfo{journal}{Phys. Rev.} \textbf{\bibinfo{volume}{D78}},
  \bibinfo{pages}{123534} (\bibinfo{year}{2008}), \eprint{0808.4085}.

\bibitem[{\citenamefont{Verde and Matarrese}(2009)}]{Verde:2009hy}
\bibinfo{author}{\bibfnamefont{L.}~\bibnamefont{Verde}} \bibnamefont{and}
  \bibinfo{author}{\bibfnamefont{S.}~\bibnamefont{Matarrese}},
  \bibinfo{journal}{Astrophys. J.} \textbf{\bibinfo{volume}{706}},
  \bibinfo{pages}{L91} (\bibinfo{year}{2009}), \eprint{0909.3224}.

\bibitem[{\citenamefont{Slosar et~al.}(2008)\citenamefont{Slosar, Hirata,
  Seljak, Ho, and Padmanabhan}}]{Slosar:2008hx}
\bibinfo{author}{\bibfnamefont{A.}~\bibnamefont{Slosar}},
  \bibinfo{author}{\bibfnamefont{C.}~\bibnamefont{Hirata}},
  \bibinfo{author}{\bibfnamefont{U.}~\bibnamefont{Seljak}},
  \bibinfo{author}{\bibfnamefont{S.}~\bibnamefont{Ho}}, \bibnamefont{and}
  \bibinfo{author}{\bibfnamefont{N.}~\bibnamefont{Padmanabhan}},
  \bibinfo{journal}{JCAP} \textbf{\bibinfo{volume}{0808}}, \bibinfo{pages}{031}
  (\bibinfo{year}{2008}), \eprint{0805.3580}.

\bibitem[{\citenamefont{Smidt et~al.}(2010)}]{Smidt:2010sv}
\bibinfo{author}{\bibfnamefont{J.}~\bibnamefont{Smidt}} \bibnamefont{et~al.}
  (\bibinfo{year}{2010}), \eprint{1001.5026}.

\bibitem[{\citenamefont{Kogo and Komatsu}(2006)}]{Kogo:2006kh}
\bibinfo{author}{\bibfnamefont{N.}~\bibnamefont{Kogo}} \bibnamefont{and}
  \bibinfo{author}{\bibfnamefont{E.}~\bibnamefont{Komatsu}},
  \bibinfo{journal}{Phys. Rev.} \textbf{\bibinfo{volume}{D73}},
  \bibinfo{pages}{083007} (\bibinfo{year}{2006}), \eprint{astro-ph/0602099}.

\bibitem[{\citenamefont{Chen et~al.}(2009)\citenamefont{Chen, Hu, Huang, Shiu,
  and Wang}}]{Chen:2009bc}
\bibinfo{author}{\bibfnamefont{X.}~\bibnamefont{Chen}},
  \bibinfo{author}{\bibfnamefont{B.}~\bibnamefont{Hu}},
  \bibinfo{author}{\bibfnamefont{M.-x.} \bibnamefont{Huang}},
  \bibinfo{author}{\bibfnamefont{G.}~\bibnamefont{Shiu}}, \bibnamefont{and}
  \bibinfo{author}{\bibfnamefont{Y.}~\bibnamefont{Wang}}
  (\bibinfo{year}{2009}), \eprint{0905.3494}.

\bibitem[{\citenamefont{Arroja et~al.}(2009)\citenamefont{Arroja, Mizuno,
  Koyama, and Tanaka}}]{Arroja:2009pd}
\bibinfo{author}{\bibfnamefont{F.}~\bibnamefont{Arroja}},
  \bibinfo{author}{\bibfnamefont{S.}~\bibnamefont{Mizuno}},
  \bibinfo{author}{\bibfnamefont{K.}~\bibnamefont{Koyama}}, \bibnamefont{and}
  \bibinfo{author}{\bibfnamefont{T.}~\bibnamefont{Tanaka}},
  \bibinfo{journal}{Phys. Rev.} \textbf{\bibinfo{volume}{D80}},
  \bibinfo{pages}{043527} (\bibinfo{year}{2009}), \eprint{0905.3641}.

\bibitem[{\citenamefont{Renaux-Petel}(2009)}]{RenauxPetel:2009sj}
\bibinfo{author}{\bibfnamefont{S.}~\bibnamefont{Renaux-Petel}},
  \bibinfo{journal}{JCAP} \textbf{\bibinfo{volume}{0910}}, \bibinfo{pages}{012}
  (\bibinfo{year}{2009}), \eprint{0907.2476}.

\bibitem[{\citenamefont{Byrnes and Tasinato}(2009)}]{Byrnes:2009qy}
\bibinfo{author}{\bibfnamefont{C.~T.} \bibnamefont{Byrnes}} \bibnamefont{and}
  \bibinfo{author}{\bibfnamefont{G.}~\bibnamefont{Tasinato}},
  \bibinfo{journal}{JCAP} \textbf{\bibinfo{volume}{0908}}, \bibinfo{pages}{016}
  (\bibinfo{year}{2009}), \eprint{0906.0767}.

\bibitem[{\citenamefont{Mizuno et~al.}(2009{\natexlab{a}})\citenamefont{Mizuno,
  Arroja, Koyama, and Tanaka}}]{Mizuno:2009cv}
\bibinfo{author}{\bibfnamefont{S.}~\bibnamefont{Mizuno}},
  \bibinfo{author}{\bibfnamefont{F.}~\bibnamefont{Arroja}},
  \bibinfo{author}{\bibfnamefont{K.}~\bibnamefont{Koyama}}, \bibnamefont{and}
  \bibinfo{author}{\bibfnamefont{T.}~\bibnamefont{Tanaka}},
  \bibinfo{journal}{Phys. Rev.} \textbf{\bibinfo{volume}{D80}},
  \bibinfo{pages}{023530} (\bibinfo{year}{2009}{\natexlab{a}}),
  \eprint{0905.4557}.

\bibitem[{\citenamefont{Mizuno et~al.}(2009{\natexlab{b}})\citenamefont{Mizuno,
  Arroja, and Koyama}}]{Mizuno:2009mv}
\bibinfo{author}{\bibfnamefont{S.}~\bibnamefont{Mizuno}},
  \bibinfo{author}{\bibfnamefont{F.}~\bibnamefont{Arroja}}, \bibnamefont{and}
  \bibinfo{author}{\bibfnamefont{K.}~\bibnamefont{Koyama}},
  \bibinfo{journal}{Phys. Rev.} \textbf{\bibinfo{volume}{D80}},
  \bibinfo{pages}{083517} (\bibinfo{year}{2009}{\natexlab{b}}),
  \eprint{0907.2439}.

\bibitem[{\citenamefont{Gao et~al.}(2009)\citenamefont{Gao, Li, and
  Lin}}]{Gao:2009at}
\bibinfo{author}{\bibfnamefont{X.}~\bibnamefont{Gao}},
  \bibinfo{author}{\bibfnamefont{M.}~\bibnamefont{Li}}, \bibnamefont{and}
  \bibinfo{author}{\bibfnamefont{C.}~\bibnamefont{Lin}},
  \bibinfo{journal}{JCAP} \textbf{\bibinfo{volume}{0911}}, \bibinfo{pages}{007}
  (\bibinfo{year}{2009}), \eprint{0906.1345}.

\bibitem[{\citenamefont{Chen and Wang}(2009)}]{Chen:2009zp}
\bibinfo{author}{\bibfnamefont{X.}~\bibnamefont{Chen}} \bibnamefont{and}
  \bibinfo{author}{\bibfnamefont{Y.}~\bibnamefont{Wang}}
  (\bibinfo{year}{2009}), \eprint{0911.3380}.

\bibitem[{\citenamefont{Chen et~al.}(2008)\citenamefont{Chen, Easther, and
  Lim}}]{Chen:2008wn}
\bibinfo{author}{\bibfnamefont{X.}~\bibnamefont{Chen}},
  \bibinfo{author}{\bibfnamefont{R.}~\bibnamefont{Easther}}, \bibnamefont{and}
  \bibinfo{author}{\bibfnamefont{E.~A.} \bibnamefont{Lim}}
  (\bibinfo{year}{2008}), \eprint{arXiv:0801.3295 [astro-ph]}.

\end{thebibliography}
\end{document}